\documentclass{article}
\usepackage[utf8]{inputenc}
\usepackage[normalem]{ulem}
\usepackage{graphicx} 
\usepackage{authblk}
\usepackage{xcolor}
\usepackage{tablefootnote}
\usepackage{cite}
\usepackage{amsmath,amssymb,amsfonts}
\usepackage{mathtools}
\usepackage{algorithmic}
\usepackage{textcomp}
\usepackage{relsize}

\usepackage[font=small,labelfont=bf]{caption}
\usepackage[square,numbers]{natbib}
\usepackage{hyperref}
\usepackage{booktabs}

\title{Assessing the impact of Open Research Information Infrastructures using NLP driven full-text Scientometrics: A case study of the LXCat open-access platform}

\author[1,2]{Kalp Pandya}
\author[1]{Khushi Shah}
\author[1, 2]{Nirmal Shah}
\author[1]{Nakshi Shah}
\author[1,2]{Bhaskar Chaudhury\thanks{Corresponding author: \texttt{bhaskar\_chaudhury@dau.ac.in}}}

\affil[1]{Group in Computational Science and HPC, DA-IICT, Dhirubhai Ambani University, Gandhinagar, India, 382007}
\affil[2]{Smart Energy Learning Center, DAU, Gandhinagar, India, 382007}

\date{ }

\begin{document}
\maketitle

\begin{abstract}
Open research information (ORI) infrastructures play a central role in shaping how scientific knowledge is produced, disseminated, validated, and reused across the research lifecycle. However, their scholarly impact is commonly assessed using citation-based and other metadata-driven indicators, which provide limited insight into how infrastructures, datasets, and software resources are actually used within scientific workflows. In this study, we present a domain-informed, natural language processing (NLP)-driven framework for full-text scientometric analysis that enables usage-aware assessment of ORI infrastructures beyond conventional citation-based measures, using the LXCat platform for low temperature plasma (LTP) research as a representative case study. The modeling of LTPs and interpretation of LTP experiments rely heavily on accurate electron and ion collision data, swarm parameters, and reaction rates associated with various gas species, much of which is hosted on LXCat, a community-driven, open-access platform central to the LTP research ecosystem.
The proposed framework integrates generic NLP techniques with domain-specific scientific knowledge to extract infrastructure-centric indicators directly from the full text of scientific publications. To investigate the scholarly impact of the LXCat platform over the past decade, we analyzed a curated corpus of approximately 400 full-text research articles citing three foundational LXCat publications. We present a comprehensive pipeline that includes data collection, cleaning, and processing, followed by the application of state-of-the-art NLP techniques informed by domain expertise. The proposed pipeline integrates chemical entity recognition, dataset and solver mention extraction, affiliation based geographic mapping and topic modeling to extract fine-grained patterns of data usage that reflect implicit research priorities, differential reliance on specific databases, evolving modes of data reuse and coupling within scientific workflows, and thematic evolution. This integrated NLP-driven approach provides new insights into how LXCat has influenced data practices and modeling priorities in LTP research.
Beyond the specific LXCat case study, the proposed framework illustrates how domain knowledge can be combined with NLP-driven full-text analysis to generate meaningful scientometric indicators of research infrastructure usage and scholarly integration that are not accessible through bibliographic metadata alone. 
This work presents a reproducible and scalable scientometric framework that has the potential to support evidence based evaluation of ORI platforms and to inform infrastructure design, governance, sustainability, and policy for future development, and advancing usage-aware scientometric analysis. To facilitate community validation and future adaptation of the framework to other domains, the complete analysis pipeline is released as open-source software.
\end{abstract}

\vspace{2pc}

\maketitle

\section{Introduction}

Open access to high-quality scientific data, including scientific databases, data repositories, software platforms, and community-driven knowledge resources has become essential for driving progress and innovation across research domains~\cite{murray2008open,thibault2023open,national2018open,knowinfra,ordplosone,openresearchbook}. Open research information (ORI) infrastructures and open research data platforms that foster data sharing, reuse, transparency, and collaboration not only accelerate innovation but also ensure reproducibility, collective advancement and reduce duplication of effort~\cite{science2004,himanen2020data,wilkinson2016fair,eresearch}. 
As these infrastructures continue to play an increasingly central role in research workflows, there is a growing need to develop methodologies that can systematically assess their scholarly usage and scientific impact. Despite the growing importance of ORI infrastructures, assessing their scientific impact remains a challenging problem. Conventional scientometric analyses predominantly rely on bibliographic metadata and indicators \cite{mingers, donthu, marzena}. 
However, these metadata-based indicators provide limited insight into how research infrastructures, datasets, software tools, and scientific databases are actually used within scientific workflows, as references to these resources are often embedded within full text of scientific publications and are not consistently represented in bibliographic metadata \cite{Mayernik,Riehl}.
Since the scientific entities that characterize infrastructure usage are inherently discipline-specific, meaningful full-text scientometric analysis requires the integration of domain knowledge.

For data intensive fields such as low temperature plasma (LTP) science, the availability of reliable datasets is particularly critical for validating plasma models, improving predictive simulations and advancing experimental diagnostics, thereby strengthening the connection between theory, simulation, and experiment~\cite{ieee2023ltp,anirudh20232022,uqltp}. LTP simulations require diverse parameters, electron and ion collision cross sections, swarm data, transport coefficients, and reaction rates, many of which are highly system dependent and difficult to obtain, and historically scattered across literature in inconsistent formats. Furthermore, data-driven approaches are increasingly enabling the discovery of new plasma applications and accelerating the integration of machine learning into plasma science workflows~\cite{kalp_eaai, trieschmann2023mlplasma, wang2020special}. 
LTP science provides an ideal domain for developing and validating full-text based scientometric methodologies because it is characterized by well-defined scientific entities, curated datasets, and widely adopted computational tools that facilitate meaningful domain-specific analysis \cite{plasmamds}. Centralized, community-curated research infrastructures play a pivotal role in supporting reproducible plasma modelling, data sharing, and interdisciplinary research within this ecosystem. Among these, the LXCat platform has emerged as a widely used open-access resource for the LTP community \cite{pitchford2017lxcat}.

The LXCat data portal (\url{www.lxcat.net}), launched in 2010, is a community-driven open platform that addresses some of these critical challenges in data-driven LTP research by providing a sustainable framework for storing, sharing, and manipulating curated data, along with online tools and validation resources essential for modeling and understanding the electron and ion components of non-equilibrium LTPs ~\cite{pancheshnyi2012lxcat, pitchford2017lxcat, carbone2021data}. Today, the LXCat project stands as one of the most prominent examples highlighting the importance of open data within the plasma physics community. 
LXCat hosts centralized databases of electron and ion scattering cross sections, swarm and transport parameters, ion-neutral interaction potentials, oscillator strengths, plasma chemistry and related datasets. It provides complete datasets required for kinetic, fluid, and hybrid plasma modeling, along with digitized swarm measurements of transport coefficients. Online tools such as BOLSIG+ solver~\cite{hagelaar2005solving} enable quick calculations of electron energy distributions, transport/rate coefficients, and comparisons of measured versus calculated data. The platform supports data validation, cross-database comparison, collaborative discussions, and offers dynamic, regularly updated datasets accessible to all with proper citation along with time machine feature which allows retrieving datasets at any given time in the past.
From the perspective of research data infrastructures, LXCat represents several important principles associated with ORI infrastructures, including open accessibility, community curation, standardized metadata, versioning, traceable data reuse and extensive community adoption. 
While LXCat is a domain-specific infrastructure rather than a comprehensive ORI ecosystem, these characteristics make it an appropriate case study for investigating infrastructure-centric scientometric methodologies.
 
 To date, more than 35 contributors from over 25 laboratories and institutions worldwide have collaboratively developed over 60 curated databases, making LXCat the most comprehensive community driven data platforms in the LTP domain ~\cite{pitchford2017lxcat}.
Over last fifteen years, LXCat has grown into an indispensable infrastructure for LTP science community, underpinning applications ranging from microelectronics fabrication, renewable energy and environmental remediation to plasma medicine, plasma assisted chemistry, green plasma technologies and space propulsion ~\cite{adamovich20222022,adamovich20172017}. By standardizing access to otherwise dispersed datasets, LXCat has addressed one of the long standing limitations of LTP research i.e. the unavailability and fragmentation of reliable input data. For several decades, data critical for LTP modeling were reported in a fragmented manner within journal articles, frequently presented only as partial plots, which made systematic reuse difficult and error-prone~\cite{national1996database}. By centralizing complete datasets, ensuring standardization and traceability through citation guidelines, LXCat has fundamentally shifted how researchers access, validate, and integrate plasma data into their workflows, thereby fostering reproducibility and enabling new interdisciplinary applications~\cite{pitchford2017lxcat}. Researchers routinely use the platform to parameterize and validate simulations, benchmark experimental measurements, and design new plasma based technologies. Its datasets have also enabled easy reproducibility of Boltzmann equation solvers, cross disciplinary innovations, and the integration of machine learning methods into plasma workflows~\cite{tejero2019lisbon,jetly2021extracting}. The breadth of LXCat’s impact suggests that it is not merely a data repository but an enabling infrastructure that has reshaped community practices around data use and sharing. In this context, LXCat serves not merely as a LTP specific resource, but as a representative case of how ORI infrastructures shape research practices, methodological choices, data reuse and collaborative patterns within a scientific community.

Despite the recognized importance of community driven open-data platforms in advancing data intensive scientific fields across disciplines, their scholarly impact remains poorly understood and rarely quantified in a systematic manner~\cite{ord}. Although several studies have evaluated domain-defining repositories through bibliometrics, citation counts, download statistics, or visibility across disciplines, such as quantifying scientific reach of the Protein Data Bank (PDB), the Sloan Digital Sky Survey (SDSS) etc., such approaches primarily capture visibility but often fail to reveal how datasets are used in practice~\cite{feng2020impact,zhang2010use}.  Similarly, despite its clear influence, only limited efforts have been made to systematically assess the scholarly impact of LXCat on LTP science and engineering. Existing evaluations rely primarily on conventional bibliometric measures, such as citation counts of its foundational publications~\cite{pancheshnyi2012lxcat, pitchford2017lxcat, carbone2021data}. While these indicators reflect academic visibility, they fail to capture how LXCat data are actually being used, whether in experimental design, computational modeling, or cross-disciplinary applications. Citations alone cannot reveal which gas species or datasets are most frequently accessed, how usage patterns differ across subfields, how databases and solvers are combined in kinetic studies, how usage patterns have evolved over time or what future directions might further strengthen the platform. This gap motivates a critical research question - how can scientific impact of research infrastructures such as LXCat be systematically evaluated, beyond citation counts, in terms of data usage, thematic trends, and evolving research practices? Addressing this question requires complementary approaches that move beyond bibliometrics to analyze full-text content and capture the nuances of data use. 
While some recent studies have highlighted the potential of full-text scientometric analysis for richer semantic characterization of scientific literature \cite{Glenisson,Riehl}, the use of domain-specific knowledge with natural language processing (NLP) to derive infrastructure-centric scientometric indicators from full text publications remains largely unexplored. Recent advances in NLP and scientometrics now make it possible to study such patterns directly from the scientific literature, offering powerful tools to assess the role of open data platforms in shaping research \cite{sciencemapping, irnlpscientometrics}.

To address this gap, we propose a domain-informed, NLP-driven framework for full-text scientometric analysis that integrates generic language processing, domain-specific scientific knowledge, and scientometric analysis to derive infrastructure-centric indicators of scholarly usage. Against this background, this work makes a unique contribution as the first systematic attempt to evaluate the scientific impact of LXCat through a large scale NLP driven analysis of nearly 400 full-text articles citing three foundational LXCat publications~\cite{pancheshnyi2012lxcat, pitchford2017lxcat, carbone2021data}. Moving beyond simple citation counts, our framework integrates domain-specific chemical entity extraction, database mention extraction, author affiliations, LTP solver mentions, and topic modeling to capture how LXCat data are used, which species and databases are most referenced, and how usage patterns have evolved over time. This approach uncovers patterns of gas usage linked to specific databases and BOLSIG+ solver, temporal shifts in modeling priorities, and country level usage and other important insights that would otherwise remain invisible. The proposed framework is designed to be transferable to other scientific domains through the incorporation of appropriate domain-specific knowledge layers. The complete source code for the implemented pipeline and metadata described in this paper are available on \href{https://github.com/nirmalshah20519/LXCat-impact-analysis}{GitHub}.\\

The key contributions of this work are as follows.\\
I) We propose a domain-informed,  full-text scientometric framework by integrating established NLP techniques with domain-specific scientific knowledge to generate infrastructure-centric scientometric indicators from full-text publications for assessing the scholarly usage and impact of research infrastructures beyond conventional metadata-based indicators. \\
II) We demonstrate the framework through a comprehensive case study of the LXCat open-access platform, illustrating how  domain specific scientific knowledge can be integrated with NLP to recover infrastructure usage patterns embedded in scientific publications.\\
III) We provide an open-source implementation of the complete full-text scientometric analysis pipeline to promote reproducibility, facilitate community validation, and support future adaptation of the framework to other scientific domains and research infrastructures. 

\section{NLP-Driven Framework for Usage-Aware Research Infrastructure Assessment}
\label{sec:nlp_ori_framework}

\begin{figure}[htbp]
    \centering
    \includegraphics[width=\linewidth]{Final_plots/LXCat_Problem_Statement.png}
    \caption{Conceptual design of the proposed domain-informed, NLP-driven framework for usage-aware scientometric assessment of ORI infrastructures. The framework integrates full-text NLP, domain-specific scientific knowledge, and scientometric analysis to transform scientific publications into infrastructure-centric indicators of scholarly usage and impact beyond conventional metadata-based bibliometric measures.}
    \label{fig:problem_statement}
\end{figure}

Conventional scientometric studies primarily assess scholarly influence through bibliographic indicators, including citation relationships, publication trends, co-authorship networks, and keyword analyses \cite{mingers, donthu, marzena}. While these indicators provide valuable measures of scholarly visibility and knowledge diffusion, they provide only limited evidence of how research infrastructures, datasets, software resources, and scientific databases are actually used within scientific practice. In contrast, the objective of the present work is to develop a systematic, scalable, usage-aware scientometric framework that derives evidence of research infrastructure usage directly from the scientific content of publications. In this work, we use the term usage-aware scientometrics to denote scientometric analyses that derive indicators from evidence of how research infrastructures are operationally used within scientific workflows, rather than solely from bibliographic metadata or citation relationships.

In this work, the term ORI infrastructure is used to refer to community-driven research infrastructures that support the collection, curation, dissemination, stewardship, citation, reuse, and exchange of research data, software, and associated scientific knowledge~\cite{science2004,knowinfra,Borgman2019DigitalArchives}. Typical characteristics associated with such infrastructures include open and persistent access, community-oriented stewardship, standardized and machine-actionable metadata, provenance and citation support, interoperability, reusability, and long-term maintainability~\cite{wilkinson2016fair,Fenner2019DataCitationRepositories}. Such ORI platforms, repositories, databases, and community-curated data portals  facilitate reproducible research, promote data sharing, and enable collaborative scientific discovery through sustained community participation ~\cite{science2004,knowinfra,wilkinson2016fair}. Rather than evaluating these infrastructures solely through citation-based visibility and bibliometric indicators~\cite{bornmann2008citation,Hicks2015LeidenManifesto,Tahamtan2016CitationFactors}, our objective is to characterize how they are integrated into scientific workflows through analysis of full-text publications. This includes investigation of which data entities are repeatedly reused, which datasets become operationally central, which computational tools are integrated into scientific workflows, or how usage patterns evolve across research directions in infrastructure-enabled science~\cite{Silvello2018DataCitation,Borgman2019DigitalArchives,Fenner2019DataCitationRepositories}. Conceptually, the proposed framework extends conventional metadata-based scientometric analysis by incorporating semantic evidence extracted from full-text publications.
This is achieved by combining generic NLP, domain-specific scientific knowledge, and scientometric analysis within a unified methodology as shown in Figure~\ref{fig:problem_statement}. Scientific publications are treated not only as citation records but also as evidence describing how research infrastructures are employed throughout the research process. 

\begin{table}[htbp]
\centering
\caption{Impact dimensions in the proposed NLP-driven framework for usage-aware assessment of ORI infrastructures, together with representative indicators and their corresponding scientometric interpretations.}
\label{tab:impact_dimensions_indicators}
\small
\renewcommand{\arraystretch}{1.35}
\begin{tabular}{p{3.3cm} p{4.7cm} p{5.3cm}}
\toprule
\textbf{Impact Dimension} &
\textbf{Representative Indicator} &
\textbf{Scientometric Interpretation} \\
\midrule

Resource adoption &
Database and dataset mentions &
Extent to which infrastructure resources are adopted by the research community \\

Computational workflow integration &
Software tool mentions (e.g., BOLSIG+) &
Integration of infrastructure resources within scientific workflows \\

Scientific application &
Domain entity (species/material/system) mentions &
Research areas supported by the infrastructure \\

Knowledge evolution &
Topic modelling &
Evolution of scientific themes enabled by the infrastructure \\

Community participation &
Author affiliations and countries &
Geographic reach and international adoption \\

Knowledge relationships &
Entity co-occurrence analysis &
Relationships among datasets, software tools, and scientific concepts \\

\bottomrule
\end{tabular}
\end{table}

Figure~\ref{fig:problem_statement}.  presents the conceptual architecture of the proposed framework that transforms unstructured scientific publications into infrastructure-centric scientometric indicators through three tightly coupled conceptual layers -\\
i) A generic information extraction layer, in which NLP techniques transform full-text publications into structured semantic representations relevant to scientific research.\\
ii) Domain knowledge integration layer, in which discipline-specific vocabularies, scientific entities, datasets, computational tools, and research workflows provide the semantic context required to interpret the extracted information. \\
iii) A scientometric indicator generation layer that transforms these enriched semantic representations into quantitative indicators of different dimensions of research infrastructure usage and scholarly impact.

The integration of these three layers enables the transition from conventional visibility-oriented scientometrics toward usage-aware scientometric analysis, where infrastructure usage is inferred from scientific content rather than bibliographic metadata alone. To clarify the relationship between extracted semantic information and infrastructure assessment, Table~\ref{tab:impact_dimensions_indicators} summarizes the principal impact dimensions considered in the proposed framework together with their corresponding indicators and scientometric interpretation. The indicators presented in this work are intended to represent proxy indicators describing complementary dimensions of research infrastructure usage rather than exhaustive measures of scientific impact. Accordingly, the proposed framework aims to complement conventional bibliometric approaches by enriching infrastructure assessment with semantically derived evidence from full-text publications.

Collectively, the proposed framework aims to shift the focus of scientometric assessment from measuring scholarly visibility through bibliographic metadata to characterizing how research infrastructures are embedded within scientific practice. By combining full-text semantic analysis with domain knowledge, the framework provides usage-aware scientometric indicators that complement conventional bibliometric measures and enable richer assessment of research infrastructure adoption and scholarly impact. 

\section{Corpus Construction and Data Extraction Pipeline}
\label{section-3}
\begin{figure}[htbp]
    \centering
    \includegraphics[width=\textwidth]{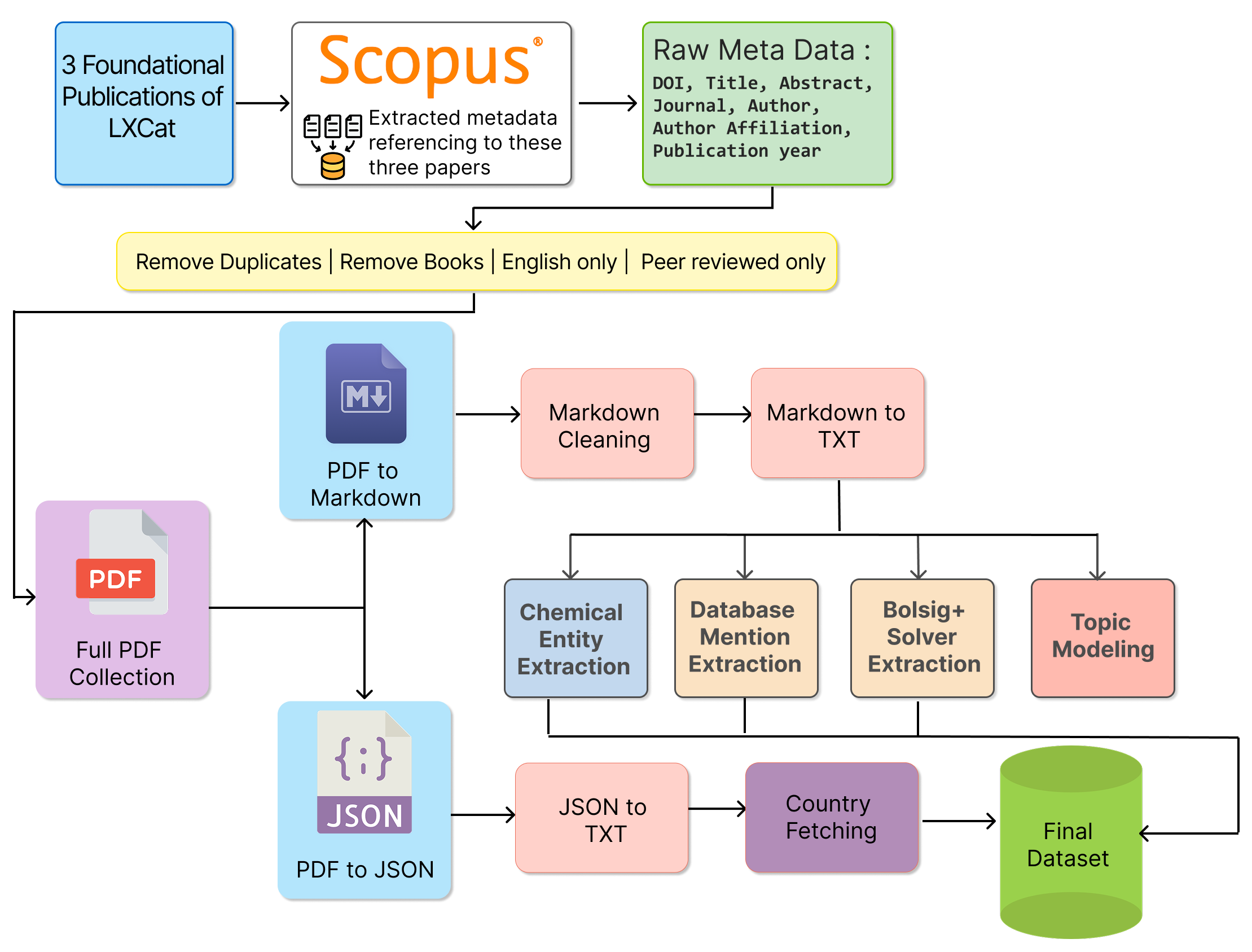}
    \caption{Overview of the data processing pipeline used to assess the scholarly impact of the LXCat platform. Starting from metadata extraction of articles citing three foundational LXCat publications~\cite{pancheshnyi2012lxcat, pitchford2017lxcat, carbone2021data}, the workflow includes deduplication, full-text PDF collection, and conversion to structured formats. Cleaned textual data is used for a suite of NLP tasks, including chemical entity recognition, dataset mention detection, BOLSIG+ solver extraction, country attribution, and topic modeling. The final curated dataset enables temporal and thematic analysis of trends in LTP research.}
    \label{fig:lxcat_datacollection}
\end{figure}

This section describes the implementation of the usage-aware scientometric framework introduced in Section~\ref{sec:nlp_ori_framework} and how it contributes to the scientometric indicators summarized in Table~\ref{tab:impact_dimensions_indicators}. Specifically, it details the construction of the literature corpus, acquisition and preprocessing of full-text publications, and the NLP-based information extraction pipeline used to derive the infrastructure-centric scientometric indicators presented later in the paper.  A reliable assessment of ORI impact requires a corpus that is methodologically well-defined and reproducible. Therefore, rather than beginning from arbitrary keyword searches, in this work the corpus is established using the three foundational publications that introduced and subsequently extended the LXCat platform ~\cite{pancheshnyi2012lxcat,pitchford2017lxcat,carbone2021data}. These publications have been selected as anchor records because they collectively document the development and evolution of the LXCat platform and provide a consistent basis for constructing the citation corpus. The overall methodology follows a multi-stage workflow, as illustrated in figure~\ref{fig:lxcat_datacollection}.
This workflow is divided into distinct phases. In the first phase, three foundational LXCat publications are selected as a reference set~\cite{pancheshnyi2012lxcat, pitchford2017lxcat, carbone2021data}, and the metadata of all citing publications are retrieved from Scopus~\cite{burnham2006scopus, boyle2006scopus}. The resulting set is refined by removing duplicates, excluding books/book chapters, and restricting the collection to english language, peer-reviewed publications only. In the second phase, full-text PDFs of the filtered records are collected and converted into structured text format to support large scale text mining. In the third phase, NLP techniques are applied to extract key information, including chemical entities (to identify gase species studied in LTP research), references to databases and BOLSIG+ solver, and country affiliations (to examine global participation and collaboration patterns). Finally, these extracted elements are integrated into a structured dataset, forming the basis for quantitative analyses of species and database usage, methodological trends, and LTP community's engagement with LXCat.

\subsection{Data Collection}
The foundational role of key LXCat publications~\cite{pancheshnyi2012lxcat,pitchford2017lxcat,carbone2021data} is well recognized within the LTP community. These works collectively outline the conceptual framework, technical development, and data centric evolution of the LXCat platform. To construct the dataset for this study, we systematically identified all peer reviewed articles citing at least one of these foundational papers using the Scopus bibliometric database, which offers broad coverage of scholarly literature and is widely adopted in scientometric analyses~\cite{burnham2006scopus, falagas2008comparison}. This citation-seeded strategy ensures that the analyzed corpus is directly connected to the scholarly community engaging with LXCat, while also providing a controlled basis for studying field growth over time. For each retrieved article, key metadata fields were extracted, including DOI, title, abstract, journal name, authors, affiliations, and year of publication. The initial dataset was further refined by excluding non-english articles, book chapters and other non-peer reviewed sources to ensure consistency and data authentication. Full-text PDFs were then collected from openly available sources, such as publisher platforms, institutional repositories, and academic databases accessible via the university. This process yielded a curated corpus of 403 full-text publications for subsequent analysis.

\subsection{PDF Conversion and Text Preparation}
\label{subsec: TXT_Preparation}

To enable large scale scientometric and natural language analysis of LXCat related literature, all full-text PDFs were first converted into structured, machine readable formats. This was achieved using the Marker framework~\cite{marker_github}, which employs a GPU accelerated, layout aware transformer model to extract reading order, text blocks, tables, captions, and metadata. Each PDF was parsed into two parallel formats: (i) a JSON structure optimized for metadata and affiliation extraction, and (ii) a Markdown (MD) version designed to support chemical species, database, and BOLSIG+ solver mention detection. To ensure completeness and robustness, a fallback recovery mechanism was implemented. PDFs were initially processed in batch mode, and any files resulting in incomplete or empty MD output were automatically reprocessed using Marker’s single file mode. The resulting JSON and MD files were stored in organized directories and constituted the foundational preprocessed dataset.

Next, we performed systematic cleaning of MD files to remove structural and non-linguistic elements that could interfere with downstream NLP. This included filtering out tables, fenced blocks, and inline mathematical expressions using targeted regular expressions. These cleaning steps helped retain only the expository textual content, thereby improving the quality and consistency of input for subsequent text analysis. Finally, cleaned MD files were converted to plain text (TXT) to produce a simplified and uniform textual representation suitable for language modeling tasks. Prior to conversion, any content under section headers containing the word ''References'' was automatically removed to exclude bibliographic material. The resulting TXT files served as the input for downstream modules such as chemical species extraction, database mention extraction, and BOLSIG\texttt{+} solver mention analysis. In parallel, the hierarchical JSON outputs generated during PDF parsing were processed to extract and normalize country names from author affiliations. Together, these TXT and JSON pathways ensured reliable extraction of domain specific entities and country level metadata for subsequent analysis.

\subsection{Derived Data Entities}

The final dataset incorporates several derived entities extracted from the processed full-text corpus, enabling quantitative and thematic analyses. These entities include chemical species, database mentions, BOLSIG+ solver mentions, topic assignments, and country affiliations, each produced through dedicated NLP pipelines described below. To ensure accuracy, all the extraction pipelines were manually validated using a small set of sample documents.

\subsubsection{Domain-specific Entity Extraction}

\begin{figure}[htbp]
    \centering
    \includegraphics[width=1\textwidth]{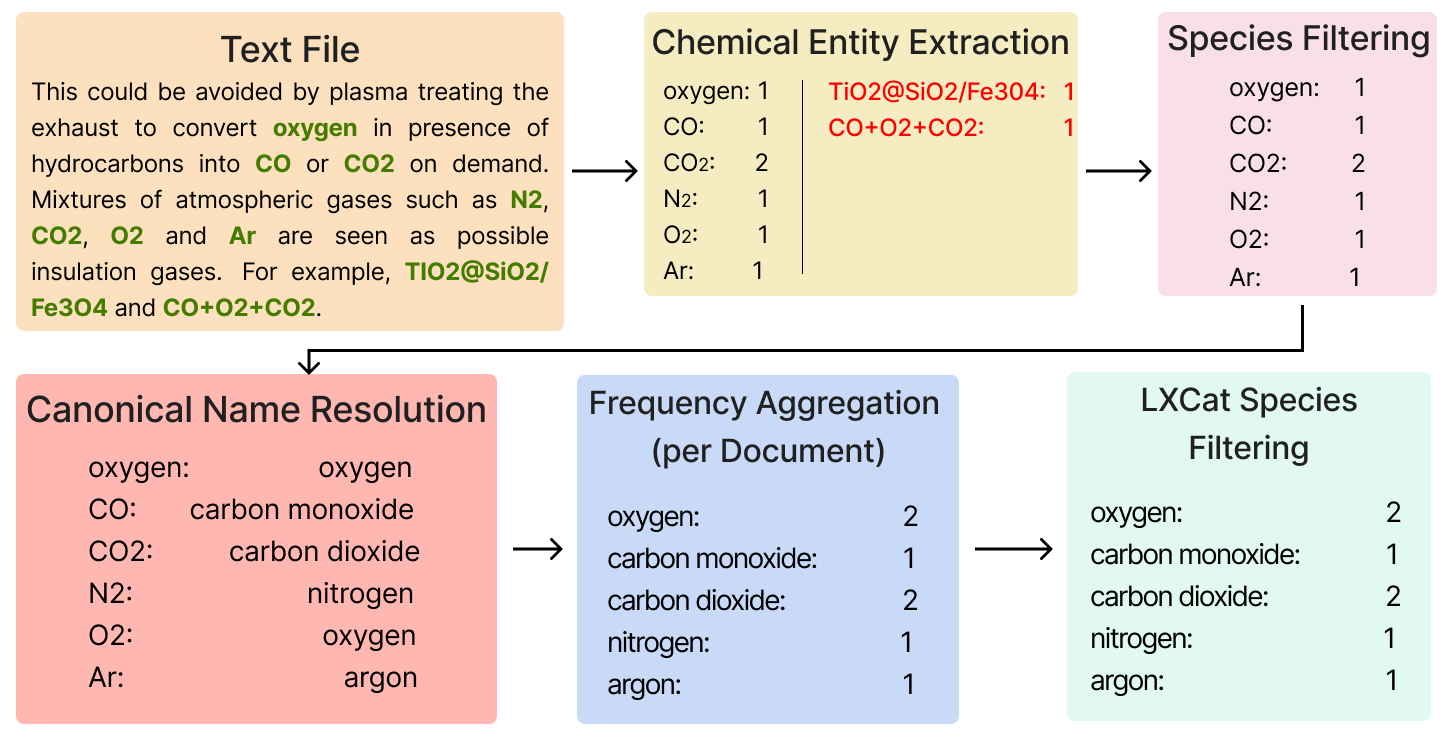}
    \caption{Overview of the chemical species extraction pipeline. Starting from raw text files, chemical entities are identified using ChemDataExtractor, followed by species-level filtering. Each entity is resolved to its canonical form via PubChem and aggregated to compute per-document frequencies. These are then filtered against a curated list of LXCat-relevant gases to produce standardized species dictionaries for downstream analysis.}
    \label{fig:gasTM}
\end{figure}

For the LXCat case study, domain-specific entity extraction corresponds to extraction of plasma species. Plasma chemical species extraction begins from the cleaned TXT files obtained from MD conversion. A multi-step pipeline systematically extracts, filters, and normalizes chemical mentions, enabling meaningful aggregation of usage patterns, as illustrated in Figure~\ref{fig:gasTM}. The pipeline processes each text file and begins by extracting chemical entities using ChemDataExtractor (CDE), a domain-specific NLP toolkit \cite{chemdataextractor}. CDE generates lists of raw chemical names, formulas, and their corresponding document level frequencies. Each entry of this list is queried against the PubChem database \cite{pubchem} to obtain its canonical name. When a direct match is not available, the first relevant synonym provided by PubChem is adopted. The resulting mapping is manually curated to remove irrelevant or misclassified terms, stored locally, and subsequently applied to resolve all chemical mentions in the corpus to standardized forms. This step ensures that synonymous variants, such as “carbon dioxide” and “CO2,” are consolidated under a unified representation. These counts are then aggregated to construct a chemical species dictionary per document. To focus on LXCat relevant species, we curate a list of ground state gases from the LXCat website and compile them into a CSV file. Elemental gases are normalized using the  \texttt{mendeleev} package \cite{Mentel_mendeleev_2021}, while compound gases and ions are resolved through PubChem \cite{pubchem}. Unmatched entries are manually verified and appended to the list. The species dictionary is then filtered against this curated set to generate \texttt{Dominant\_species\_dictionary}, representing chemical species explicitly present in LXCat.

\subsubsection{Research Infrastructure Resource Extraction}

\begin{figure}[htbp]
    \centering
    \includegraphics[width=1.0\textwidth]{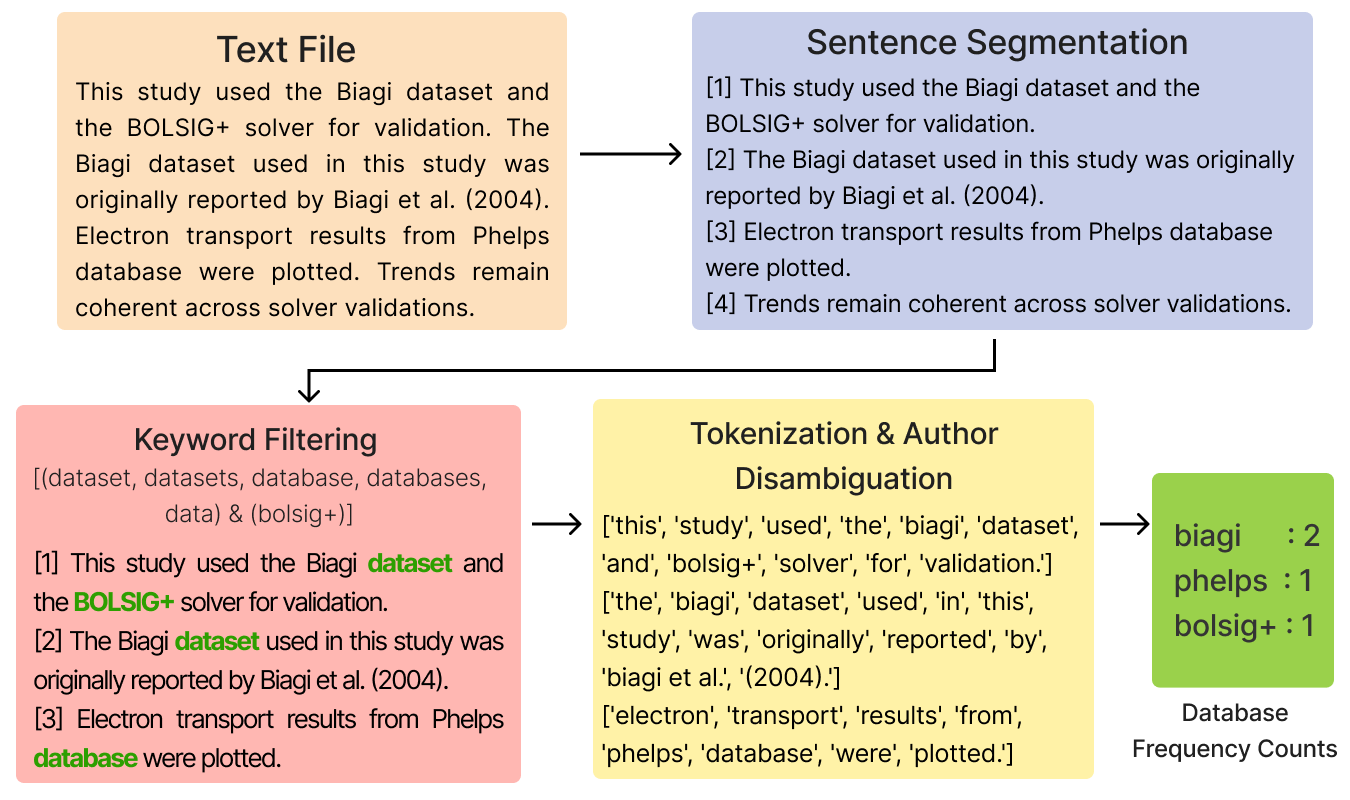}
    \caption{Overview of the database mention extraction pipeline. Starting from plain text files, the workflow applies sentence segmentation, followed by keyword filtering to isolate data relevant sentences. Tokens are then processed through tokenization and author name disambiguation, after which database names are identified and aggregated into document-level frequency counts.}
    \label{fig:db_tm}
\end{figure}
In this LXCat case study, research infrastructure resource extraction corresponds to extraction of databases.
Database mention extraction operates on cleaned TXT files obtained from MD files. A systematic pipeline, shown in Figure~\ref{fig:db_tm}, identifies, extracts, and aggregates references to LXCat databases through sentence segmentation, keyword filtering, tokenization, and author name disambiguation. The resulting normalized mentions are compiled into document-level frequency counts for downstream analysis.

The pipeline processes each TXT file by first segmenting multi-line text paragraphs using a custom and robust sentence splitter, that preserves abbreviations, decimal numbers, acronyms, and other scientific notations to avoid false boundaries. After applying the keyword filtering module, only sentences containing user-defined data-related keywords are retained for further analysis. Further on, the selected sentences are tokenized such that the author names are grouped into separate author tokens (e.g., “Biagi et al.”) and handled independently, ensuring that they are not confused with the database names appearing elsewhere in the sentence. All remaining tokens are matched against a curated vocabulary of known LxCat databases, with strict exact token comparison to prevent accidental matches with unrelated words. Detected database mentions are counted, normalized across version variants, and aggregated into a document-level dictionary containing the databases, their frequencies, and the extracted keyword-relevant sentences.

\subsubsection{Computational Tool Extraction}

BOLSIG+ solver (computational tool) extraction is performed on the cleaned TXT files obtained from MD files. Since BOLSIG+ is widely used to calculate electron transport coefficients and rate coefficients from cross-section data, we extended our text mining approach to specifically identify and quantify references to BOLSIG+ within the corpus. The BOLSIG+ extraction pipeline automatically identifies and analyzes all meaningful occurrences of the solver across scientific text files by applying a robust sentence splitter that preserves abbreviations, decimal numbers, acronyms, and other scientific notations to avoid false boundaries, and filtering only those sentences that explicitly contain the solver keyword. 

After this, a custom tokenizer is applied to correctly handle bracketed terms and technical expressions, enabling accurate detection of both direct and parenthetical mentions. All versioned or modified solver names are normalized into a single base form to produce reliable frequency counts. For each document, the pipeline outputs clean list of sentences containing BOLSIG+ mentions and the total number of mentions, providing both qualitative context and quantitative insight into BOLSIG+ usage across the literature.

\subsubsection{Country Fetching}

The country fetching pipeline is designed specifically to identify author affiliations by scanning only the first two pages of each scientific document where affiliations are conventionally located. The system processes structured JSON files, recursively collecting text-bearing blocks such as paragraphs, list items, and footnotes from pages 0 and 1. All HTML markups are removed, and broken lines are reconstructed into clean sentences to ensure accurate detection. A comprehensive country mapping module, built from ISO-standard country~\cite{iso3166} names plus common affiliation variants (e.g., “USA,” “UK,” “Czech Republic”), is then applied to normalize and extract country names even when written in different formats. The pipeline aggregates all detected country mentions corresponding to author affiliations and stores them in a clean CSV, providing a reliable automated method for affiliation based country extraction across large corpora of research papers.

\section{Final Dataset Structure}
Following corpus construction and full-text preprocessing described in Section \ref{section-3}, the extracted information is organized into a structured dataset that supports both conventional bibliometric characterization and infrastructure-centric full-text scientometric analysis as summarized in Table \ref{tab:final_dataset_structure}. The resulting dataset contains multiple complementary levels of information corresponding to the different stages of the proposed framework.

At the first level, bibliographic metadata describe the overall characteristics of the corpus, including publication year, journal name, DOI information, citation metadata, authors, institutional affiliations and abstract. These data provide the baseline scientometric context for analysing the temporal growth, publication venues, and geographical distribution of the LXCat-related literature.
At the second level, full-text NLP processing generates domain-specific semantic information from the body of each publication. This includes identified plasma species, references to LXCat databases and related scientific resources, mentions of computational tools such as the BOLSIG+ solver, and country information derived from author affiliations. Collectively, these extracted entities constitute the semantic layer used to characterize different dimensions of infrastructure usage. The resulting structured dataset therefore combines conventional bibliographic metadata with semantically enriched full-text information, providing the foundation for the infrastructure-centric scientometric analyses presented in the following sections.

\begin{table}[htbp]
\centering
\caption{Structure of the final LXCat-related publication dataset, comprising bibliographic metadata and derived full-text NLP fields for scientometric analysis of species, database usage, BOLSIG+ mentions, and geographic distribution.}
\begin{tabular}{p{3cm} p{10cm}}
\hline
\textbf{Field Name} & \textbf{Purpose / Utility} \\
\hline
\multicolumn{2}{l}{\textbf{Metadata Fields}} \\
\hline
Original DOI & The three foundational LXCat publications used~\cite{pancheshnyi2012lxcat,pitchford2017lxcat,carbone2021data} \\
Citing DOI & Research papers that cite these foundational publications \\
Title & Identifies research focus \\
Published Year & Temporal analysis \\
Journal Name & Source validation \\
Authors & Attribution and collaboration mapping \\
Author Affiliations & Institutional and country mapping \\
Abstract & Thematic and keyword analysis \\
\hline
\multicolumn{2}{l}{\textbf{Derived Data Fields}} \\
\hline
\textbf{Chemical Species Dictionary} & \textbf{Dictionary of chemical species from LXCat, along with their frequencies of occurrence} \\
\textbf{Database Dictionary} & \textbf{Dictionary of scientific databases mentioned in the paper, along with their citation frequencies} \\
\textbf{BOLSIG+ Counts} & \textbf{Total number of times BOLSIG+ is mentioned in the paper} \\
\textbf{Country} & \textbf{The countries of the institutional affiliations of all authors of the paper.} \\
\hline
\end{tabular}
\label{tab:final_dataset_structure}
\end{table}

We provide dataset cardinality statistics to highlight its breadth and uniqueness. The corpus includes 403 unique citing papers (data captured as of May 2025) from 120 journals, representing authors from 72 countries. The derived fields further capture 232 unique species, 35 distinct databases, and more than 600 references to BOLSIG+. To obtain these, we processed approximately 120K sentences to systematically extract gas mentions, database mentions, and BOLSIG+ counts. These coverage metrics demonstrate the dataset’s comprehensiveness and reinforce its value for large scale scientometric and thematic analysis.

\section{Topic Modeling}

\begin{figure}[htbp]
    \centering
    \includegraphics[width=1 \textwidth]{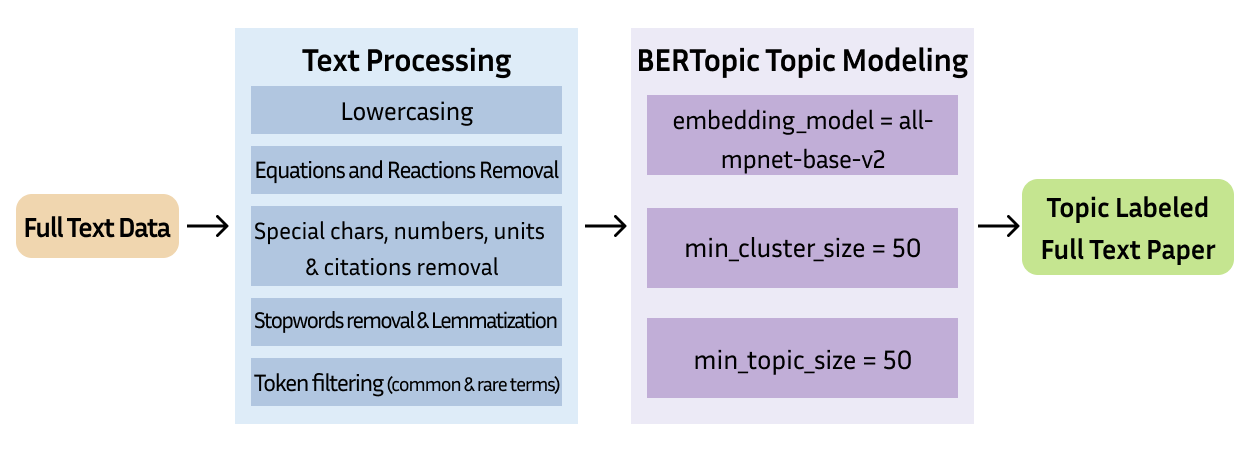}
    \caption{Overview of the topic modeling pipeline. Full-text documents undergo multiple preprocessing steps, including lowercasing, removal of equations, units, citations, and non-informative tokens, followed by lemmatization and frequency-based token filtering. The cleaned corpus is then passed to BERTopic. The output consists of full-text documents labeled with their most representative topic.}
    \label{text_processing}
\end{figure}

To explore latent themes within the corpus, we employ topic modeling on the preprocessed TXT files from subsection~\ref{subsec: TXT_Preparation}, following the workflow illustrated in figure~\ref{text_processing}. The preprocessing pipeline ensures a clean and coherent foundation by converting all text to lowercase, removing equations, chemical reactions, and citation markers, applying regex-based cleaning for special characters, units, and single letter tokens, and further refining the corpus through stopword removal and lemmatization. In addition, rare tokens that appeared only a few times across the corpus, as well as overly frequent tokens that lacked discriminative power, are filtered out to improve topic coherence and reduce noise.

On this refined corpus, we employ BERTopic~\cite{grootendorst2022bertopic}, which leverages transformer-based embeddings to capture contextual semantics of words and phrases. The embeddings are clustered using density-based algorithms (HDBSCAN)~\cite{campello2013density}, and the resulting groups are represented through class-based TF-IDF (c-TF-IDF)~\cite{salton1988term}, enabling the extraction of coherent and interpretable topics. As illustrated in Figure~\ref{text_processing}, we provide details of the key hyper-parameters used in the BERTopic model-namely, the embedding model ('all-mpnet-base-v2'), 'min\_cluster\_size', and 'min\_topic\_size'. These parameters were fine-tuned iteratively by validating the output topics with domain experts and adjusting the settings to enhance interpretability and relevance. Unlike traditional probabilistic approaches such as LDA~\cite{blei2003latent}, BERTopic provides improved topic coherence by integrating contextualized language representations. The resulting topics highlight domain specific research directions, allowing us to trace thematic structures, methodological trends, and their evolution within the LXCat related literature.

\section{Results}

\begin{figure}[htbp]
    \centering
    \includegraphics[width=\textwidth]{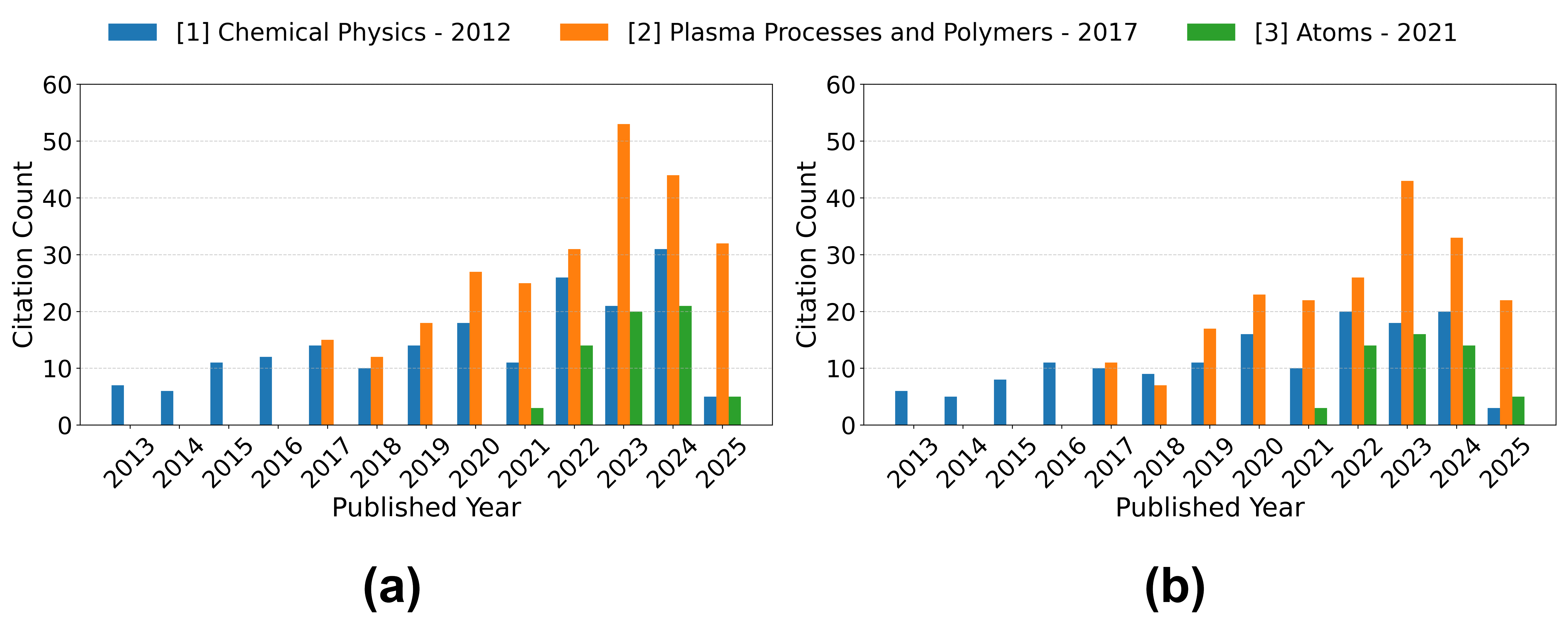}
    \caption{Annual citation trends for the three foundational LXCat publications. (a) Full Scopus citation counts from $\sim$500 publications. (b) Citation counts for $\sim$400 articles with full-text access used in this study for NLP based analysis.}
    \label{fig:a1}
\end{figure}

\begin{figure}[htbp]
    \centering
    \includegraphics[width=\textwidth]{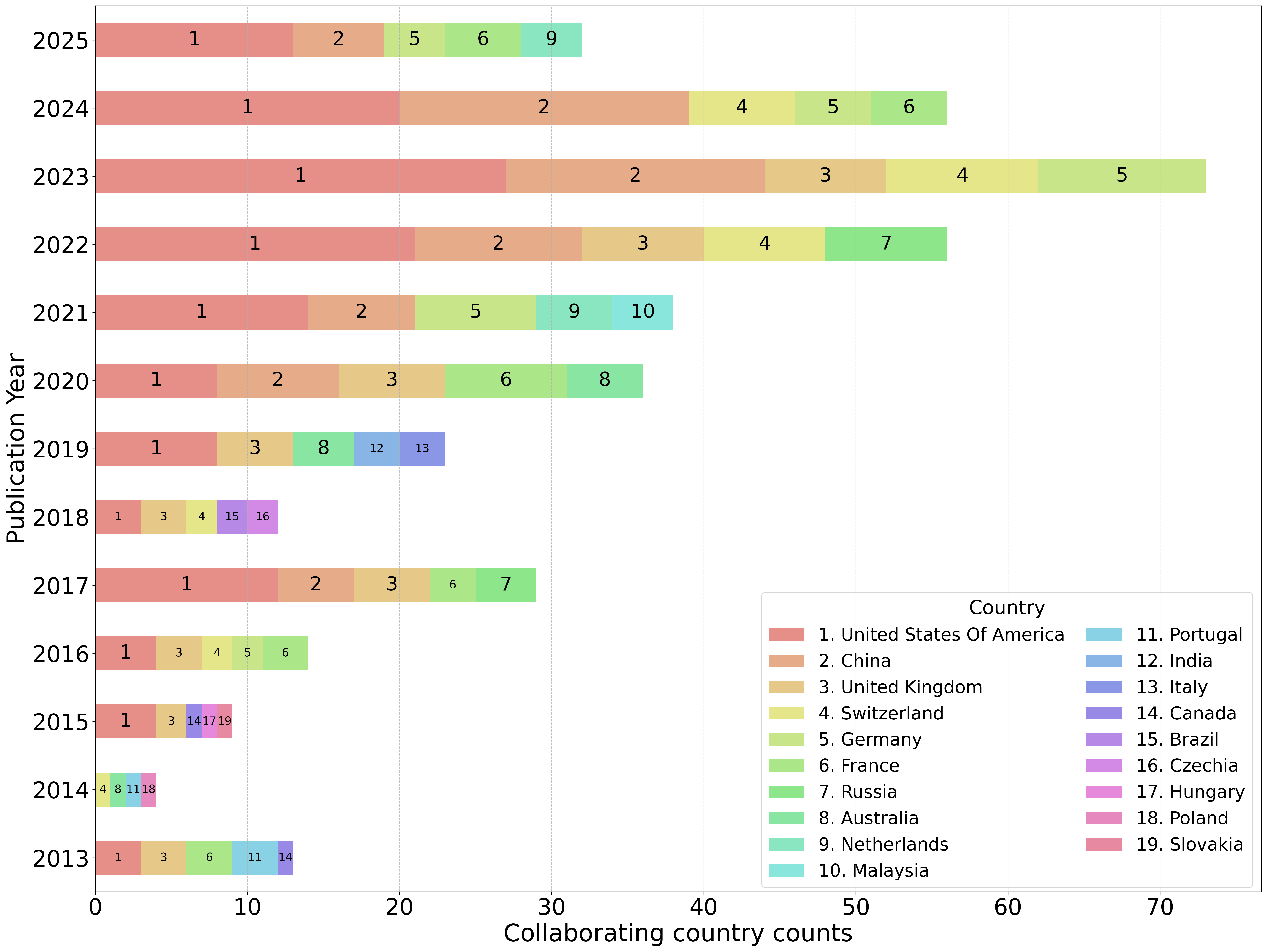}
    \caption{Top five contributing countries per year for LXCat-citing publications (2013-2025), based on author affiliations. Color-coded numeric labels correspond to the ranking of different countries in terms of total number of LXCat-citing publications attributed to each country across the full period  (e.g., United States~(1), China~(2), United Kingdom~(3)).}
    \label{fig:country}
\end{figure}

Building on the full-text analysis pipeline and curated corpus described in the preceding sections, in this section we present results that characterize how the LXCat open data platform is referenced and integrated within the LTP research literature over the past decade. The temporal patterns reported in this section should be interpreted in the context of the accessible full-text corpus analyzed in this study as summarized in Section~\ref{section-3} and Figure~\ref{fig:a1}. While the foundational LXCat publications have been cited by a larger body of literature, full-text availability constraints necessarily limit our analysis to a subset of citing articles for which complete texts could be obtained and processed. Consequently, the observed temporal trends do not represent absolute citation volumes, but rather the evolution of observable 
LXCat engagement within accessible scholarly literature. The goal of this analysis is to reveal the broader patterns of LXCat usage, and its scholarly and thematic impact on the LTP community as reflected in scholarly discourse. Figures ~\ref{fig:a1} and \ref{fig:country} provide an initial, high-level view of the temporal evolution and structural breadth of LXCat engagement, establishing the empirical context for the more detailed gas species and database usage oriented analyses, co-occurrence mapping, temporal trends, and topic modeling that follow. 
Collectively, these distributions serve as first-order indicators of ORI infrastructure diffusion, community reliance, and maturation, illustrating how a domain-specific open data platform evolves into a trusted and routinely integrated ORI infrastructure spanning journals, institutions, and geographic regions.

Figure~\ref{fig:a1} presents the annual citation distribution for the three foundational LXCat publications, comparing the complete Scopus dataset Figure~\ref{fig:a1}a with the curated full-text subset analyzed in this study Figure~\ref{fig:a1}b. Notably, the \textit{Plasma Processes and Polymers} 2017 publication~\cite{pitchford2017lxcat} exhibits a huge citation increase in recent years, whereas the \textit{Chemical Physics} 2012 publication~\cite{pancheshnyi2012lxcat} has contributed a steady baseline of citations over a longer period. Together, these trajectories indicate both sustained engagement within the LXCat citation profile. The similarity between the two sub-plots suggests that the full-text subset provides a representative view of broader usage patterns. From an ORI perspective, this reflects LXCat's growing integration into data-driven research practices and its maturation as a reusable infrastructure within the LTP community. All subsequent analyses in this study therefore operate on the full-text accessible subset.

Figure~\ref{fig:country} summarizes the temporal evolution (across the entire 2013-2025 period) of country-level participation in LXCat-citing publications, restricted to the top five contributing countries per year using a horizontal stacked bar chart. The United States consistently appears as the leading contributor, with China, the United Kingdom, and several European countries showing increasing participation over time. The appearance of countries such as Australia, India, Brazil, and Malaysia in later years reflects a widening international user base. This diversification signals the geographic diffusion of ORI infrastructure enabled research practices and the expanding role of LXCat as a globally accessed data resource. These observations underscore the relevance of geographic coverage as an indicator of inclusive and equitable ORI adoption.

\begin{figure}[htbp]
    \centering
    \includegraphics[width=1\textwidth]{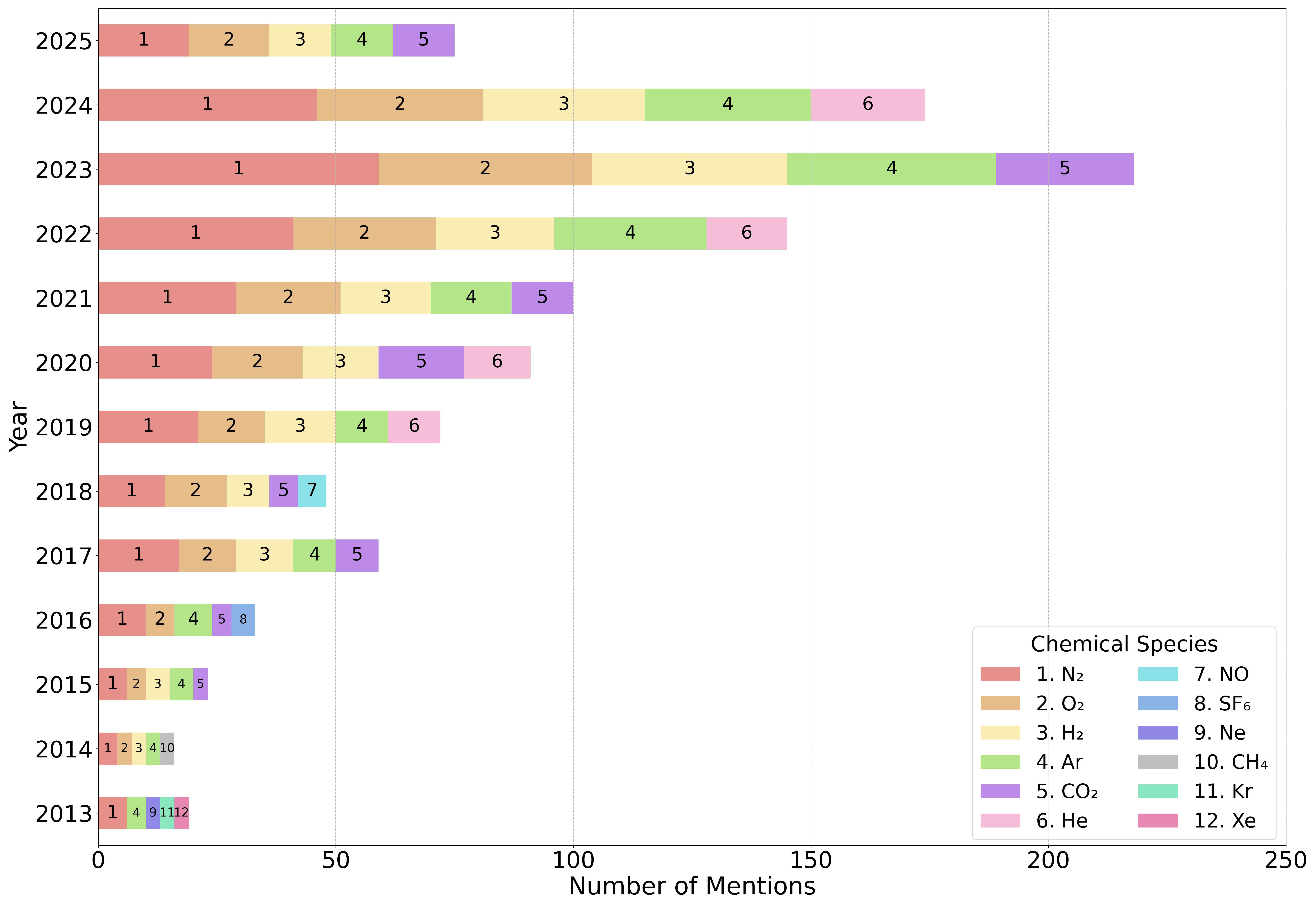}
    \caption{Temporal trends in mentions of the top five chemical species across LXCat-citing publications (2013-2025). Mentions are extracted at the sentence level; a single species may be referenced multiple times within a single article. Color-coded numeric labels denote cumulative mention counts across the full period and are ordered in descending rank (e.g., N$_2$~(1), O$_2$~(2), H$_2$~(3)).}
    \label{fig:gas_temporal_trend}
\end{figure}

Figure~\ref{fig:gas_temporal_trend} illustrates the evolution of the most frequently referenced chemical species within the full-text corpus of LXCat citing publications. Nitrogen (N$_2$) and oxygen (O$_2$) consistently constitute the highest number of mentions, reflecting their centrality in gas discharge physics and LTP modeling. Hydrogen (H$_2$), argon (Ar), and carbon dioxide (CO$_2$) also exhibit sustained representation, indicative of their relevance across combustion, gas conversion, and environmental plasma applications. Species with lower absolute counts, including helium (He), nitric oxide (NO), and sulfur hexafluoride (SF$_6$), potentially point to more specialized or emerging research areas. The numeric labels correspond to the cumulative total mentions of each species across the 2013-2025 period and define their descending rank within the stacked distributions. These fine-grained temporal patterns highlight the utility of full-text extraction in revealing shifts in data demand and modeling priorities that are not visible through citation metadata alone, underscoring how open data platforms such as LXCat shape and support evolving research priorities.

\begin{figure}[htbp]
    \centering
    \includegraphics[width=1\textwidth]{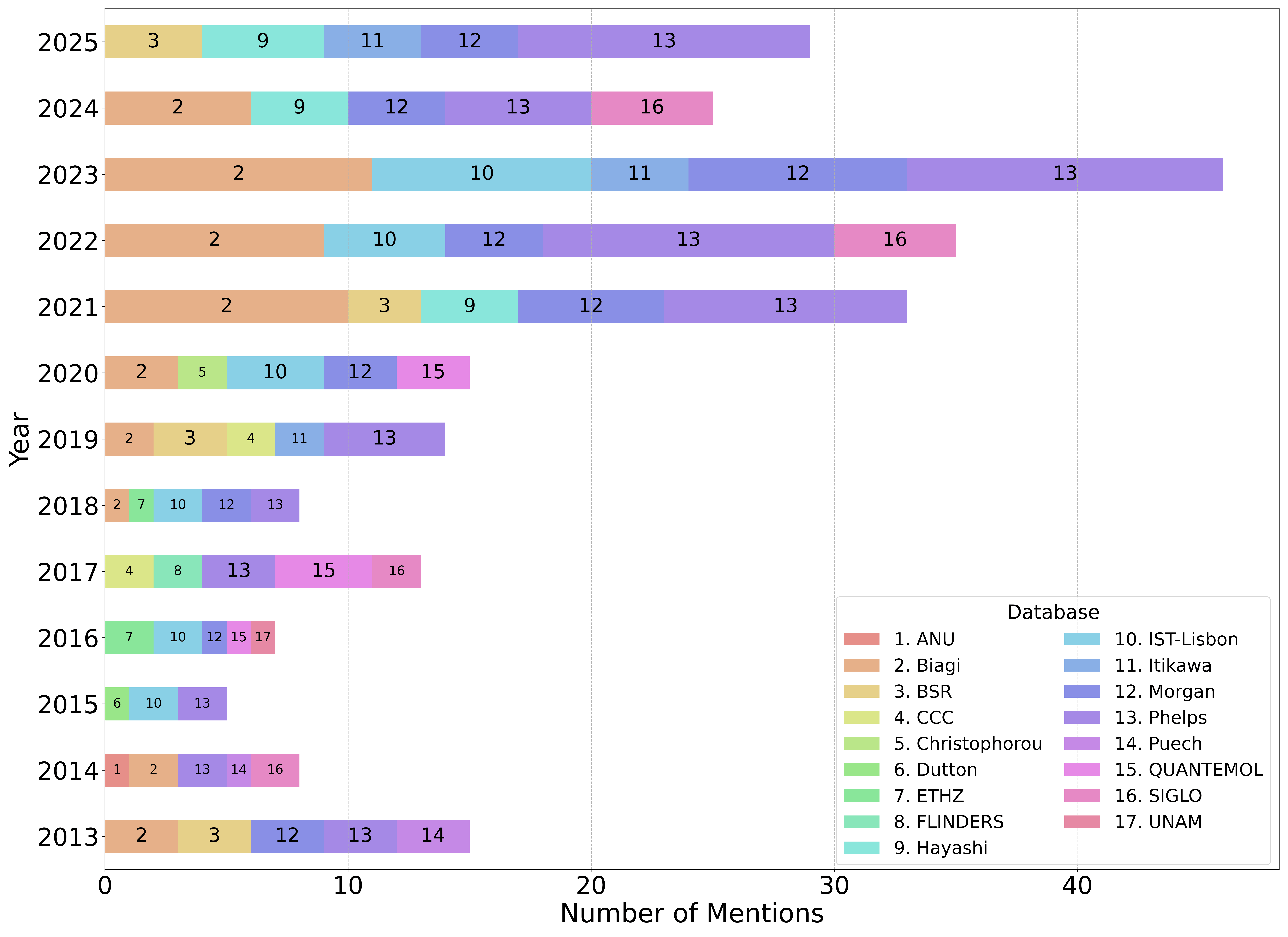}
    \caption{Temporal trends in mentions of the top five LXCat databases per year (2013-2025), extracted from sentence-level text acrosss the curated full-text corpus. Mentions reflect total occurrences of each database name in individual papers, with multiple mentions possible per article. Color-coded numeric labels correspond to database identifiers ordered alphabetically (e.g., ANU~(1), Biagi~(2), BSR~(3)).}
    \label{fig:dbs_temporal_trend}
\end{figure}

Figure~\ref{fig:dbs_temporal_trend} presents the annual frequency of sentence-level mentions of LXCat-hosted databases in publications citing the three foundational LXCat papers. The analysis captures multiple mentions per article, enabling fine-grained insight into database usage patterns over time. Phelps (13), Biagi (2), Morgan (12), and IST-Lisbon (10) appear consistently across years, indicating sustained relevance within the LTP modeling community. Other databases, such as Itikawa (11), Hayashi (9), BSR (3) and SIGLO (16), show episodic peaks, suggesting more targeted or problem-specific applications. The numeric labels in the bars follow an alphabetical ordering of database names and do not reflect usage rank. Overall, the distribution highlights a shift from reliance on a small number of well-established databases toward broader engagement with a diverse set of data resources. From an ORI perspective, this trend aligns with evolving community norms emphasizing cross-validation, modular reuse, and infrastructure-mediated reproducibility in data-intensive plasma science research.

\begin{figure}[htbp]
    \centering
    \includegraphics[width=\textwidth]{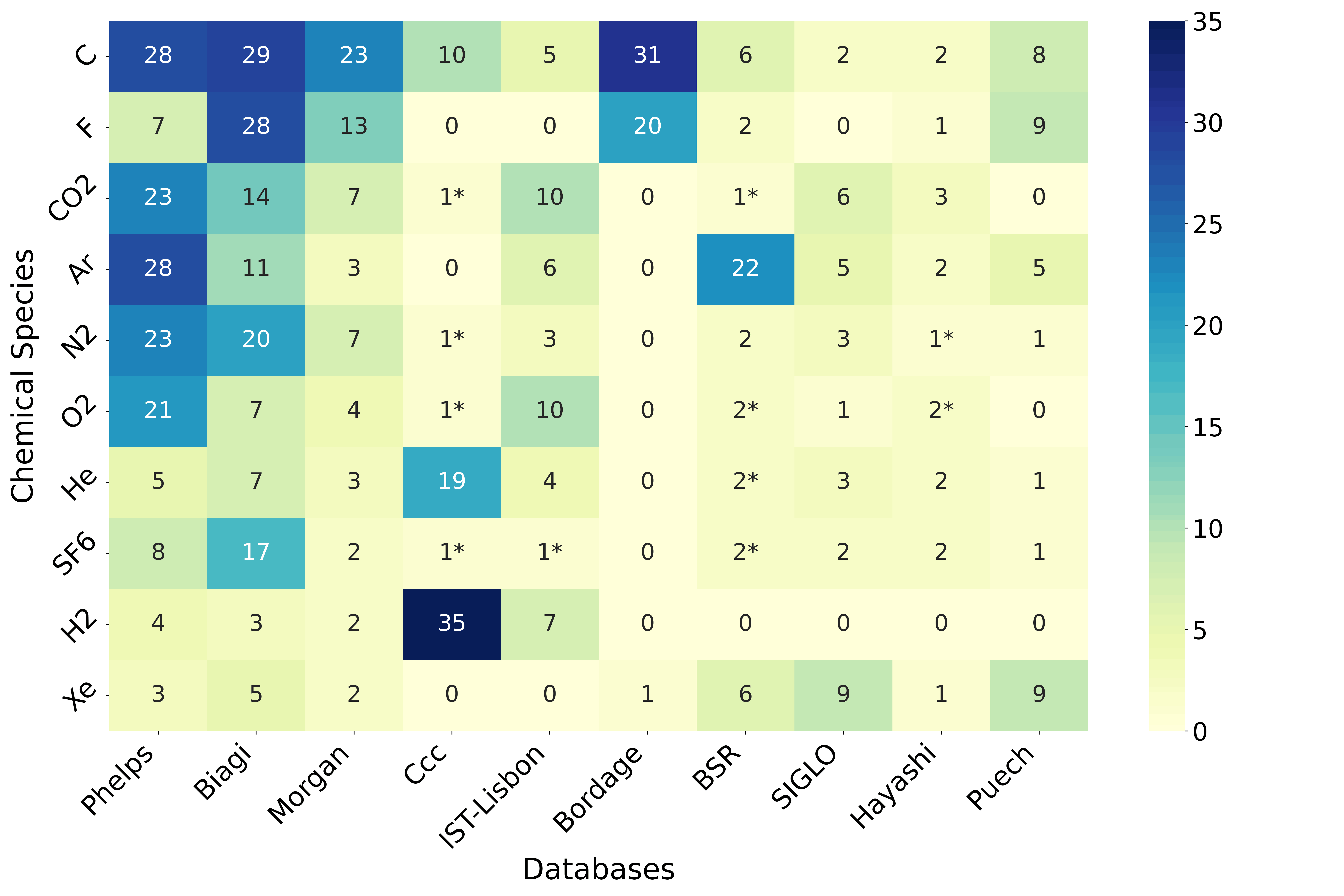}
    \caption{Co-occurrence heatmap showing sentence-level associations between chemical species and LXCat databases across the full-text corpus. Each cell denotes the frequency with which a species and a database are mentioned within the same sentence. Values marked with '*' indicate cases where the species is not explicitly contained within the corresponding database but appears in contextual proximity, suggesting indirect relevance.}
    \label{fig:gas_db_co}
\end{figure}

Figure~\ref{fig:gas_db_co} presents sentence-level co-occurrence patterns between frequently referenced chemical species and LXCat-hosted databases. Prominent associations between chemical species (e.g. N$_2$, O$_2$, CO$_2$, Ar, C and F) and multiple databases highlight the breadth of electron collision data and swarm parameters accessed from LXCat in support of LTP modeling. The presence of contextual associations marked with '*' indicates instances in which database mentions appear alongside species not explicitly included in that database, reflecting indirect coupling through modeling workflows, solver usage, or comparative discussion. Such co-occurrence structure provides evidence of how open data resources are integrated within scientific discourse, revealing potential coupling patterns between domain entities and data infrastructures that are not identifiable from citation metadata alone.

\begin{figure}[htbp]
    \centering
    \includegraphics[width=\textwidth]{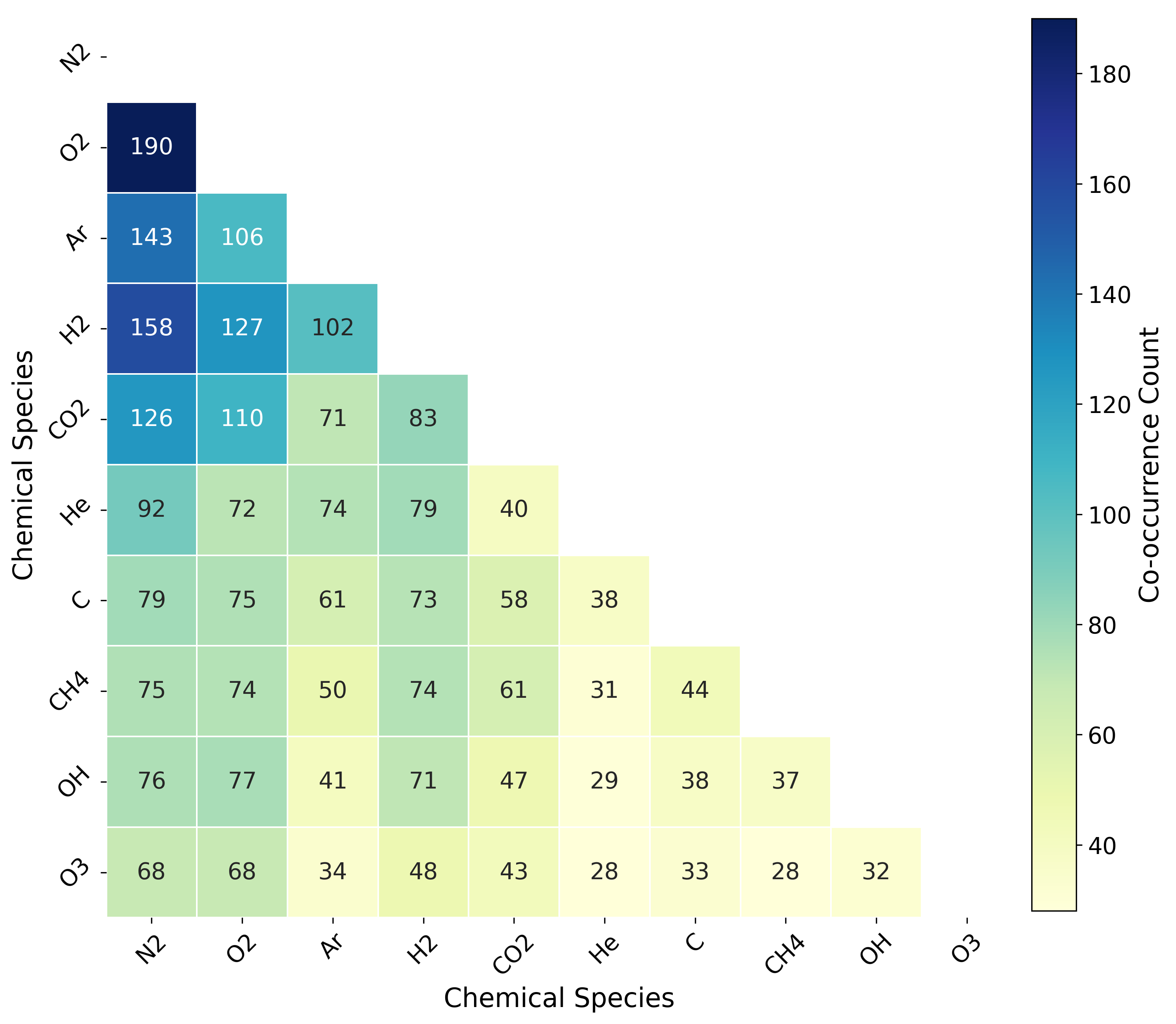}
    \caption{Chemical species co-occurrence heatmap derived from document-level analysis, indicating how often pairs of gases are mentioned together in the same publication. Higher co-occurrence values suggest shared modeling or experimental contexts.}
    \label{fig:gas_gas_heatmap}
\end{figure}

Figure~\ref{fig:gas_gas_heatmap} shows the document-level co-occurrence frequencies between frequently mentioned gas species in the full-text corpus. Nitrogen (N$_2$) and oxygen (O$_2$) form the most prominent pair, followed by frequent co-mentions involving argon (Ar), hydrogen (H$_2$), and carbon dioxide (CO$_2$). These patterns reflect common groupings of species used in LTP modeling and suggest shared physical contexts such as gas mixtures, or target species in discharge processes. Lighter co-occurrence among species such as methane (CH$_4$), hydroxyl radical (OH), and ozone (O$_3$) points to more domain-specific modeling scenarios. From an ORI standpoint, these results highlight the value of entity-level full-text extraction for identifying implicit relationships among datasets and modeling priorities that support integrative use of platforms like LXCat in complex multi-species research environments.

\begin{figure}[htbp]
    \centering
    \includegraphics[width=\textwidth]{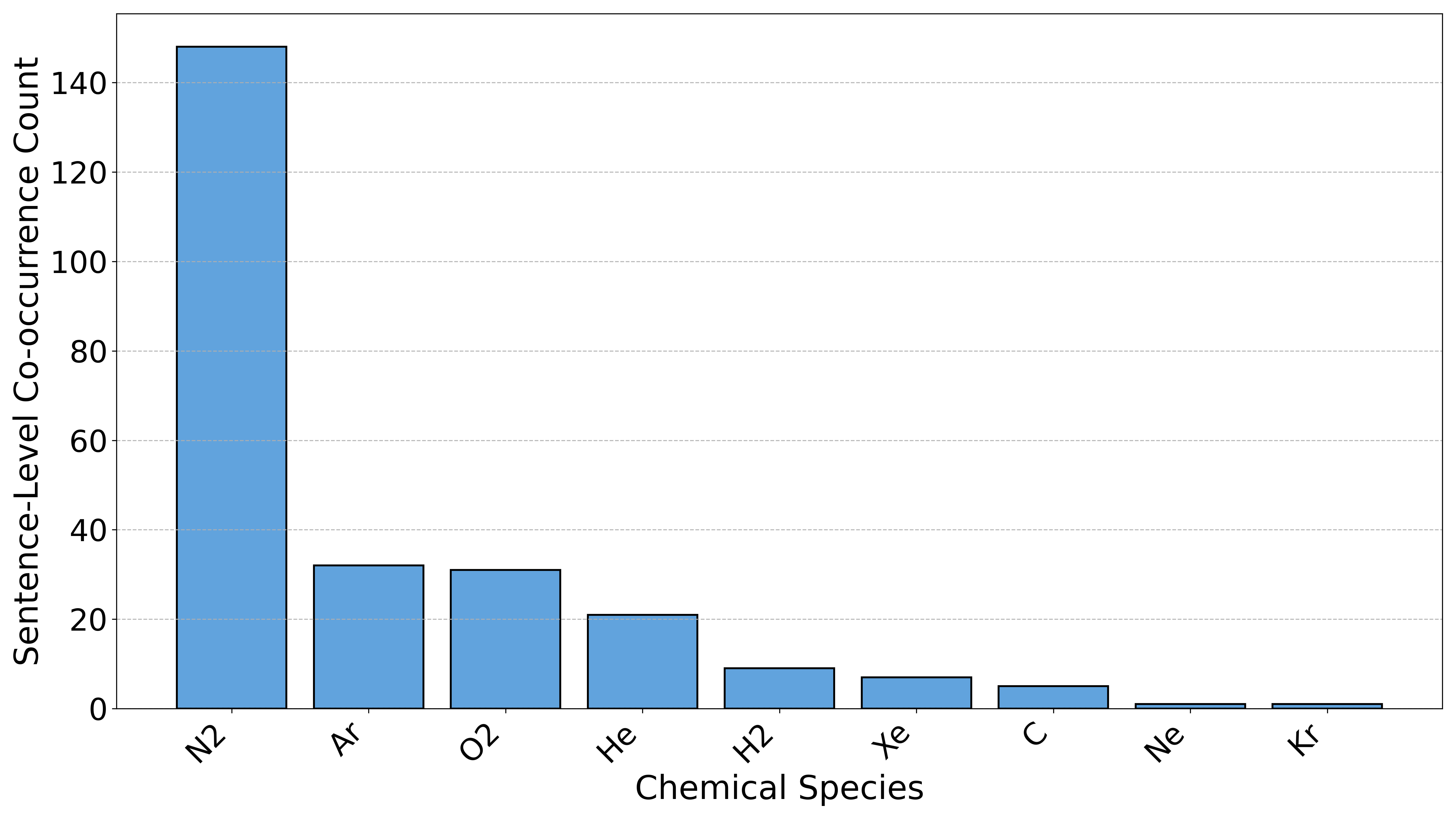}
    \caption{Top ten chemical species that co-occur with the BOLSIG+ solver at the sentence level within the analyzed corpus. Species may appear multiple times within the same article, and counts reflect aggregated sentence-level co-occurrences.}
    \label{fig:top_gases_bolsig}
\end{figure}

Figure~\ref{fig:top_gases_bolsig} summarizes the species that most frequently co-occur with the BOLSIG+ solver within the corpus. Nitrogen (N$_2$) dominates the distribution, followed by argon (Ar) and oxygen (O$_2$), reflecting their central role in electron kinetics, swarm calculations, and discharge modeling tasks for which BOLSIG+ is commonly employed. Species such as Helium (He) and hydrogen (H$_2$) appear with moderate frequency, consistent with their relevance in atmospheric pressure and combustion related plasma applications. Other species such as xenon (Xe), atomic carbon (C), neon (Ne), and krypton (Kr) occur far less frequently, indicating narrower domain usage tied to specific experimental or modeling contexts. These co-occurrence patterns highlight how solver usage provides insight into the operational layer of ORI enabled research activity. Rather than indicating mere visibility, the associations reveal the types of species for which LXCat linked solvers are applied in practice, illustrating how open data (collision cross sections) and computational tools (BOLSIG+) become coupled within data-driven plasma modeling workflows.

\begin{figure}[htbp]
    \centering
    \includegraphics[width=\textwidth]{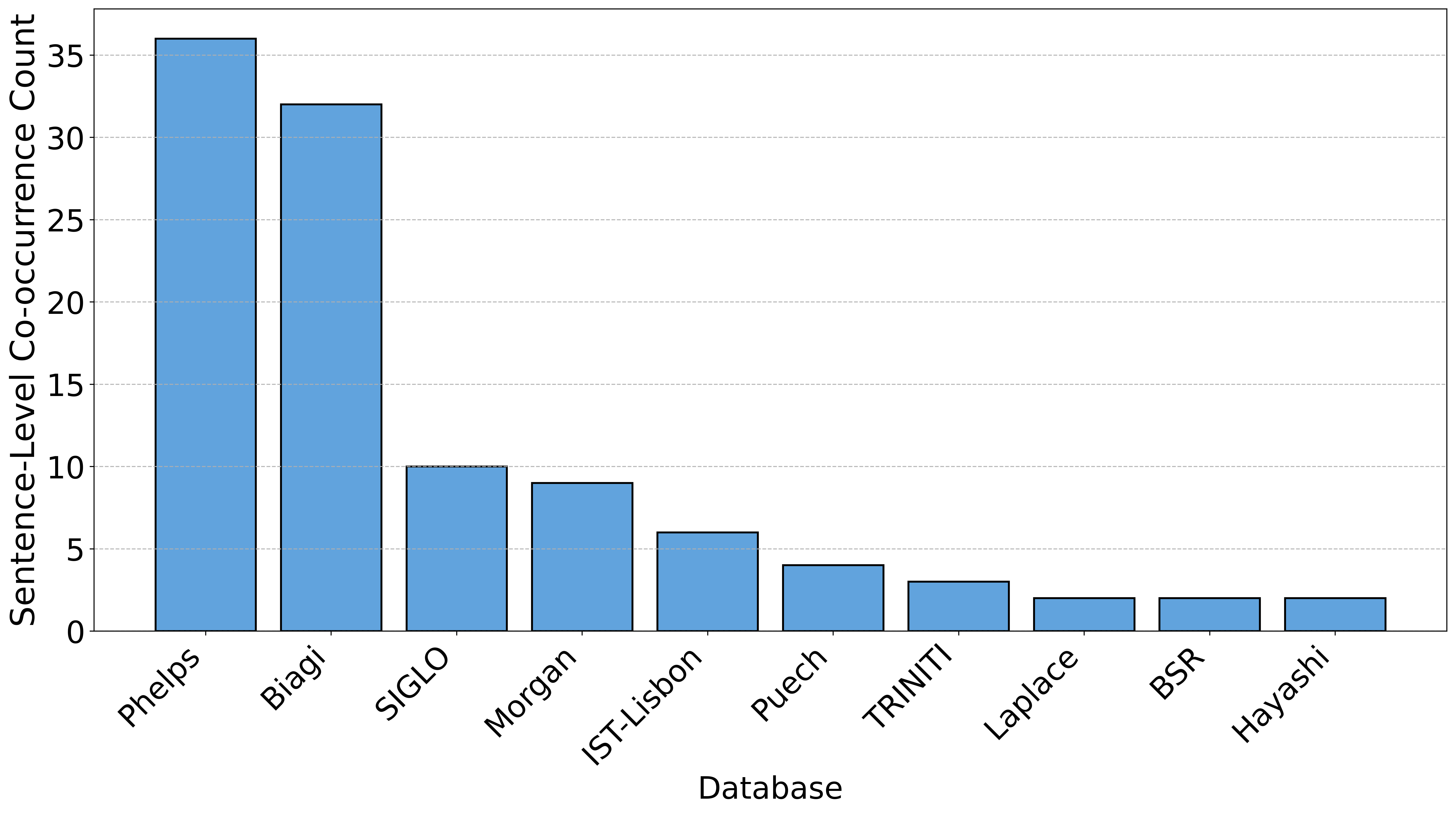}
    \caption{Top ten LXCat databases that co-occur most frequently with the BOLSIG+ solver at the sentence level. These co-occurrences reflect instances where the solver and a database are referenced in proximity, indicating practical coupling of data and modeling tools in simulation workflows.}
    \label{fig:top_db_bolsig}
\end{figure}

Figure~\ref{fig:top_db_bolsig} shows the most frequently co-mentioned LXCat databases in BOLSIG+-related sentences across the corpus. The Phelps and Biagi databases dominate this distribution, indicating their central role in providing cross-section datasets used for electron transport modeling. The proximity of solver and database mentions offers insight into the functional coupling of computational tools and curated datasets, underscoring the role of LXCat not only as a data repository, but as a platform that supports modeling interoperability, reuse, and tool-data integration. This sentence level co-occurrence analysis thus reveals active open data platform use beyond citation metrics, aligning with usage-aware approaches to ORI impact assessment.

\begin{figure}[htbp]
    \centering
    \includegraphics[width=\textwidth]{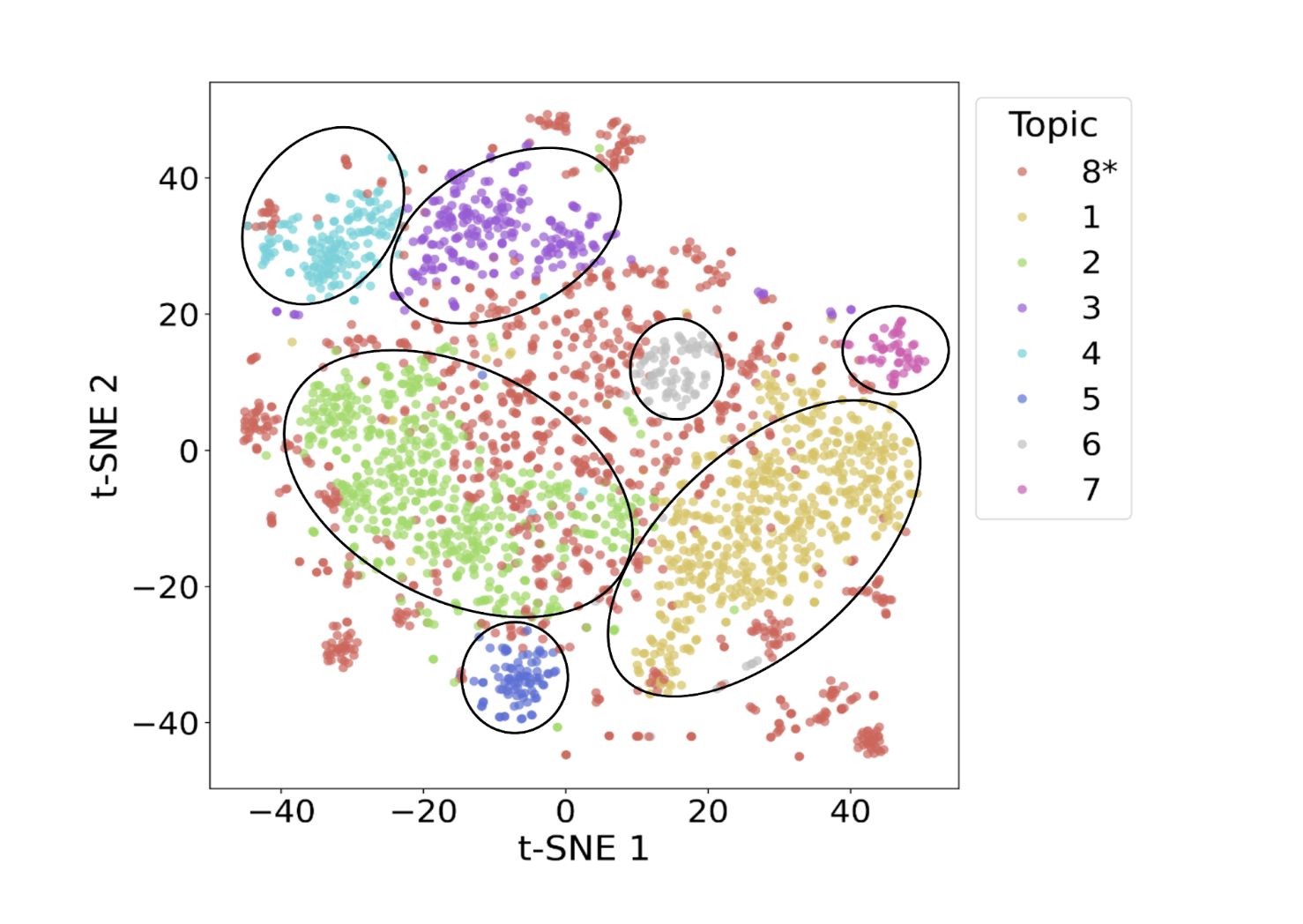}
    \caption{t-SNE visualization of BERTopic-derived thematic clusters from the LXCat-citing corpus. Topic 8* reflects documents with diffuse or overlapping themes, while the remaining topics correspond to coherent and recurring research areas.}
    \label{tsne}
\end{figure}

\begin{table}[htbp]
\centering
\caption{Thematic clusters or topics extracted by BERTopic.}
\begin{tabular}{p{1.3cm} p{6.5cm} p{5cm}}
\hline
\textbf{Topic No.} & \textbf{Top Words} & \textbf{Topic Name} \\
\hline
1 & cross, section, electron, energy, scattering, data, ev, coefficient, ionization, collision & Electron Collision \& Cross Sections \\
2 & electron, discharge, density, plasma, field, streamer, electric, figure, simulation, voltage & Discharges \& Plasma Dynamics \\
3 & co2, plasma, conversion, catalyst, reaction, co, energy, gas, temperature, vibrational & CO\textsubscript{2} Conversion \& Catalysis \\
4 & ignition, reaction, flame, plasma, specie, discharge, pulse, energy, temperature, radical & Plasma Combustion \& Ignition \\
5 & thruster, plasma, magnetic, electron, ion, simulation, plume, case, thrust, domain & Plasma Thrusters \& Space Propulsion \\
6 & data, plasma, model, research, example, physic, community, standard, science, used & Plasma Data, Modeling \& Tools \\
7 & system, radiation, process, condensed, multiscale, matter, md, particle, molecular, approach & Condensed Matter \& Multiscale Sims \\
8* & plasma, electron, model, ion, density, energy, field, reaction, gas, specie & Miscellaneous \\
\hline
\end{tabular}
\label{tab:topics}
\end{table}

The topic modeling analysis using BERTopic on full-text documents revealed eight coherent thematic
clusters within the corpus of LXCat-related publications, as summarized in Table~\ref{tab:topics}. These topics capture both the scientific focus areas and the methodological frameworks
prevalent in the literature. Figure~\ref{tsne} presents a two-dimensional t-SNE projection of the BERTopic generated clusters. The visual structure reveals semantically coherent groupings, with Topics 1 and 2 focusing on electron collisions and discharge dynamics occupying the densest and most expansive regions. Application focused topics, including CO\textsubscript{2} conversion (Topic 3), plasma combustion (Topic 4), and space propulsion (Topic 5), appear as discrete and bounded clusters. Thematic areas around modeling tools and multiscale methods (Topics 6 and 7) are also well-separated, while Topic 8* comprises more diffusely located points, capturing texts that do not align strongly with dominant topical axes. From an ORI standpoint, these results validate the breadth of LXCat's engagement across diverse subdomains and illustrate the role of ORI infrastructure in sustaining thematically differentiated yet interoperable research directions.

Taken together, the analyses presented in this section provide a multidimensional view of LXCat as an evolving ORI infrastructure. Our results position LXCat not just as a data repository, but as a dynamic infrastructure actively shaping knowledge production across the LTP research lifecycle. Importantly, these findings were made possible by a full-text, NLP driven approach that captures actual usage contexts offering a more granular and usage aware alternative to traditional citation metrics. 

\section{Discussion}
 The results presented in this study indicate that LXCat functions as an important ORI infrastructure that is increasingly getting embedded within data-driven LTP research and validation practices. The prominence of a small number of trusted databases, alongside the increasing prevalence of multi-database usage, suggests an evolution of community norms toward data reliability,  cross-validation, reproducibility and emergence of shared standards. The observed co-occurrence between specific chemical species, curated databases, and tools such as BOLSIG+ further illustrates how ORI infrastructures shape not only what data are accessed, but how they are operationalized within scientific research. In this sense, LXCat usage shows how domain-specific open data platforms can evolve into epistemic infrastructures that embed data, tools, and practices within a shared research ecosystem. This integrative role aligns closely with broader ORI objectives of promoting transparency, accountability, and reproducibility across the research lifecycle. 

 The LTP community's reliance on a small set of well-established databases (e.g., Phelps, Biagi) underscores the need for continued curation, expansion, and validation to support emerging research priorities. At the same time, the increasing incidence of multi-database usage and solver-data integration highlights the potential for advancing cross-validation, and FAIR-aligned practices within the field.  The results presented in this paper offer key insights into several pathways for further strengthening LXCat's role as a sustainable research infrastructure. Future research could focus on (i) expanding coverage to underrepresented species and reactions (ii) developing quantitative indicators of dataset quality and reliability, (iii) tracing data reuse pathways from LXCat into experimental design and industrial applications, (iv) extending NLP based impact studies to capture software/tool integration beyond BOLSIG+, (v) fostering community wide protocols for establishing standards for metadata, dataset annotation, benchmarking and uncertainty quantification, (vi) supporting integration with machine learning workflows and (vii) to develop interactive dashboards for realtime monitoring of species/database trends. Addressing these directions would possibly enhance LXCat’s value as a community resource and will further strengthen LXCat’s role as a FAIR-aligned open science infrastructure for the global LTP community. In summary, this work highlights LXCat’s transformation from a data repository into an enabling infrastructure that has redefined community practices around data sharing and reuse in plasma science. 

Importantly, the contributions of this study extend beyond a domain-specific impact assessment of LXCat. By applying full-text NLP driven scientometric analysis to a large corpus of publications citing foundational LXCat papers, we uncover patterns of data reuse, methodological dependence, and thematic evolution that remain largely invisible to conventional citation based indicators.
From a broader ORI standpoint, these findings demonstrate how open, community-driven data platforms function as active components of knowledge production rather than static information storage portals. The framework presented here provides a means to interrogate this active role empirically, offering a more fine-grained understanding of ORI impact. Although LXCat and the LTP community serve as the empirical focus of this study, the proposed framework is designed to be adaptable across scientific domains through appropriate domain-specific customization as detailed in Section \ref{sec:generalizability_domain_agnostic_pipeline}. Core components of the pipeline including gas species entity extraction, dataset mention disambiguation, software and tool detection, author affiliation retrieval, and topic modeling can be readily adapted to other ORI contexts. For example, chemical species extraction can be mapped to genes, proteins, or materials in other disciplines, solver identification corresponds to detecting domain-specific software tools, and database mention analysis generalizes naturally to repositories such as the Protein Data Bank, GenBank, or large-scale Earth system data portals. This cross-domain mapping, summarized in Table~\ref{tab:lxcat_ori_mapping}, highlights the broad applicability of the approach.

A limitation of the present study arises from the construction of the citation-seeded corpus. The analysis includes publications that explicitly cite one or more of the three foundational LXCat papers and therefore does not capture studies that use LXCat resources without formal citation. Conversely, not every citation necessarily indicates direct operational use of the platform, as some publications may cite LXCat only for contextual or illustrative purposes. Accordingly, the indicators presented in this work should be interpreted as evidence of infrastructure usage within the population of publications that explicitly acknowledge LXCat, rather than as an exhaustive measure of total platform usage.

\begin{table}[htbp]
\centering
\caption{Mapping of LXCat-specific components to broader ORI functions}
\label{tab:lxcat_ori_mapping}
\begin{tabular}{p{3.8cm} p{4.3cm} p{5.3cm}}
\toprule
\textbf{Component used in this study} \newline (LXCat specific) & \textbf{Function in this Study} & \textbf{Corresponding ORI Component} \\
\midrule
Chemical species extraction & Identifies gases used in LTP research & Domain entities applicable in other areas of research \\
Database mention extraction & Tracks data usage patterns & Repository/database/dataset mentions in other OpenData Projects \\
BOLSIG+ solver detection & Captures LTP modeling tool and open data coupling & Software/workflow identification such as simulation codes, analysis pipelines etc. \\
Topic modeling & Reveals thematic research trajectories in LTP & Research direction mapping and cross-disciplinary ORI studies \\
Country affiliation analysis & Measures LXCat global participation and impact & Equity and geographic inclusion for ORI policy, governance and future development \\
Full-text NLP pipeline & Captures implicit data use & Usage-aware scientometrics beyond citation-based evaluation \\
\bottomrule
\end{tabular}
\end{table}
Beyond methodological generalizability, the findings carry potential implications for the design, governance, and sustainability of ORI infrastructures. From a policy perspective, the proposed methodology offers developers, funders, and policymakers a data-driven mechanism to monitor adoption of open data portals, identify gaps in coverage, and provide implicit signals of community priorities. Such evidence based insights can complement existing assessment frameworks and support more responsive and sustainable evolution of ORI systems. 

\section{Generalizability and Cross-Domain Adaptation of the Proposed Framework}
\label{sec:generalizability_domain_agnostic_pipeline}
Although the proposed framework is demonstrated using the LXCat platform, its underlying methodology is intended to be applicable to a much broader class of research infrastructures. The objective of this work is to demonstrate a general framework for domain-informed, NLP-driven full-text scientometric analysis that can be adapted to assess the scholarly usage and impact of research infrastructures across scientific disciplines.
The proposed framework adopts a modular architecture that separates generic analytical components from the domain knowledge layer required for scientific interpretation. Generic components, including corpus construction, full-text processing, NLP, co-occurrence analysis, topic modelling, and visualization, operate independently of the scientific domain. In contrast, the domain knowledge layer encapsulates discipline-specific scientific entities, controlled vocabularies, repositories, datasets, software tools, and scientific resources that provide the semantic context necessary for interpreting the extracted information. Consequently, adaptation of the framework to a different scientific domain primarily involves replacing the domain knowledge layer while preserving the overall analytical workflow. Our modular software design, available at \href{https://github.com/nirmalshah20519/LXCat-impact-analysis}{GitHub repository}, simplifies customization for alternative research infrastructures. In future implementations, the Scopus-dependent retrieval stage can also be replaced with open scholarly infrastructures such as the OpenAIRE Graph, OpenCitations or OpenAlex, thereby supporting a more openly accessible and reproducible pipeline.

\begin{table}[htbp]
\centering
\caption{Framework components and corresponding domain-specific adaptations in the proposed NLP-driven assessment pipeline.}
\label{tab:framework_domain_adaptation}
\small
\renewcommand{\arraystretch}{1.35}
\begin{tabular}{p{4.5cm} p{8.5cm}}
\toprule
\textbf{Framework Component} &
\textbf{Domain-Specific Adaptation} \\
\midrule

Corpus construction &
Select domain-relevant seed publications for corpus identification and retrieval \\

PDF processing &
Generic module; no domain-specific adaptation required \\

NLP preprocessing &
Generic module; no domain-specific adaptation required \\

Entity extraction &
Replace the entity vocabulary with terminology specific to the target domain \\

Repository extraction &
Replace the repository vocabulary with domain-relevant repository names and resources \\

Software extraction &
Replace the software vocabulary with tools and computational resources relevant to the target domain \\

Topic modelling &
Retrain the topic model on the target-domain corpus \\

Co-occurrence analysis &
Generic module; no domain-specific adaptation required \\

Scientometric indicators &
Interpret indicators in accordance with domain-specific expertise and research context \\

\bottomrule
\end{tabular}
\end{table}

Table \ref{tab:framework_domain_adaptation} summarizes the principal components of the proposed framework and distinguishes those that are domain-independent from those requiring domain-specific adaptation. As shown, most stages of the workflow remain unchanged across application domains, while only the semantic resources used for information extraction and interpretation require customization. Therefore, in the framework design, the generic NLP techniques provide the computational capability for extracting structured information from scientific publications, whereas domain expertise provides the scientific semantics necessary to transform these extracted signals into meaningful scientometric indicators.  Unlike conventional bibliometric analyses, which predominantly operate on structured publication metadata and citation relationships, the proposed usage-aware full-text scientometric analysis seeks to recover evidence embedded within the scientific content of publications. Meaningful interpretation of such content inevitably depends on disciplinary knowledge, since identical textual patterns may have fundamentally different scientific meanings across research fields. Accordingly, developing an appropriate domain knowledge layer represents a necessary step for obtaining scientifically meaningful and interpretable indicators of infrastructure usage. We believe that this methodological paradigm provides a practical foundation for future investigations of research infrastructures across diverse scientific domains and contributes towards the development of richer, usage-aware approaches for assessing the scholarly impact of open science and research information infrastructures.

\section{Conclusion}
This work presents a NLP-driven and domain-informed framework for full-text scientometric analysis to assess the scholarly usage and impact of ORI infrastructures beyond conventional bibliometric indicators. While traditional scientometric analyses predominantly rely on bibliographic metadata, the proposed framework demonstrates how evidence embedded within the full text of scientific publications can be transformed into meaningful indicators of infrastructure usage. By integrating generic NLP techniques with domain-specific scientific knowledge, the framework enables the extraction and interpretation of entities, datasets, software tools, computational workflows, and thematic relationships that are typically inaccessible through conventional bibliometric analyses. In this sense, the proposed methodology complements existing scientometric approaches by providing a richer, usage-aware perspective on the scientific role of research infrastructures.

The framework has been demonstrated through a comprehensive case study of the LXCat platform (an open-access, web-based platform designed to store, curate, display, and distribute data both for modeling and for interpretation of LTP experiments), one of the most widely used community-driven data infrastructures in LTP science. Analysis of nearly 400 full-text publications citing the three foundational LXCat articles revealed how the platform has become deeply embedded within plasma research workflows. Beyond confirming the broad scientific visibility of LXCat, the proposed methodology identified patterns of database usage, chemical species associated with different databases, coupling between LXCat datasets and the BOLSIG+ computational solver, temporal evolution of research themes, geographical distribution of scientific activity, and emerging application areas.  Our results also  demonstrate how LXCat has evolved from a data repository into a community-wide infrastructure that informs, supports, and shapes modeling practices, data workflows, and thematic research directions.  By systematically quantifying its scientific impact, we provide both a reflective assessment of its achievements, translational impact of open plasma data and a roadmap for its future evolution. These findings illustrate how full-text analysis can uncover evidence of infrastructure usage and scientific integration that cannot be inferred from citation counts or other bibliographic indicators alone.

A central contribution of this work is the demonstration that domain knowledge constitutes an essential component of meaningful full-text scientometric analysis. Unlike conventional metadata-based approaches, interpreting scientific content requires understanding the discipline-specific entities, vocabularies, datasets, software tools, and research workflows through which infrastructures contribute to scientific practice. Accordingly, the proposed framework combines generic NLP with a domain knowledge layer that provides the semantic context necessary for transforming textual evidence into scientifically meaningful usage indicators. Although the present implementation has been developed for the plasma science community, the underlying analytical workflow is intentionally modular and can be adapted to other research domains by incorporating the corresponding domain-specific knowledge. We therefore view this work as an initial proof-of-concept towards domain-informed, usage-aware scientometric analysis, where infrastructure assessment is derived from scientific content rather than bibliographic metadata alone.

The proposed framework also contributes to the broader goals of open science by providing a transparent and reproducible methodology for evaluating community-driven research infrastructures. The release of the complete analysis pipeline as open-source software facilitates independent validation, community participation, and future adaptation to other scientific domains.  Future work will focus on extending the framework toward more context-aware and actionable infrastructure assessment. In particular, we plan to develop context-aware citation and mention analysis to distinguish direct dataset reuse, methodological dependence, contextual discussion, and illustrative references. Additional directions include developing quantitative indicators of dataset quality and reliability. Collectively, these extensions would strengthen the interpretability, reproducibility, and practical utility of usage-aware scientometric assessment for research infrastructures. We believe that the present study represents an important first step towards next-generation scientometric methodologies that combine full-text analytics, natural language processing, and domain expertise to provide richer and more nuanced assessments of how research infrastructures enable scientific discovery.

\section*{Code Availability}
The brief details of this project, with code availibility can be found on our web page \href{https://sites.google.com/dau.ac.in/gics-codes/gics-scientometrics}{GICS Codes}. The website also consists \href{https://github.com/nirmalshah20519/LXCat-impact-analysis}{GitHub repository}, which includes modules for metadata extraction, full-text conversion (PDF to JSON/MD/TXT), chemical entity extraction, database and solver mention detection, topic modeling using BERTopic, and country-level author affiliation analysis, as outlined in figure~\ref{fig:lxcat_datacollection}. Documentation and example scripts are provided to facilitate reproducibility and adaptation to related datasets. For further details or collaboration inquiries, researchers are encouraged to contact the corresponding author. Contributions to this open-source project are welcome.

\section*{Acknowledgment}

The authors gratefully acknowledge Prof.\ Leanne Pitchford (LAPLACE, CNRS \& Université de Toulouse, France) for her constructive feedback and valuable LXCat related discussions during the development of this work. Authors also acknowledge Agam Shah (Georgia Institute of Technology, USA) for  valuable NLP related suggestions. In addition, the authors acknowledge the Information and Library Network (INFLIBNET) Centre, Gandhinagar, India, for providing access to the Scopus database, which was essential in the data collection phase of this research. The authors sincerely thank the anonymous reviewers for their insightful and constructive feedback, which has greatly enhanced the clarity and overall quality of the article. KP, NS and BC acknowledge
BSES Rajdhani Power Limited and BSES Yamuna Power Limited for the CSR grants to carry out the work at the Smart Energy Learning Centre (SELC), Dhirubhai Ambani University (DAU), Gandhinagar, India.

\bibliographystyle{abbrv}
\bibliography{references}

\end{document}